\documentclass[10pt]{wlscirep2}
\usepackage{fontspec}
\usepackage{polyglossia}
\usepackage{csquotes}
\usepackage{placeins}
\usepackage{amsmath}

\usepackage{textcomp}
\usepackage{amsmath}
\usepackage{array, multirow, bigdelim, makecell, booktabs} 

\usepackage{libertinus}

\setdefaultlanguage{english} 
\setotherlanguage{russian}

\usepackage{bm}
\usepackage{braket}
\usepackage{booktabs}
\usepackage{amsmath}
\usepackage{multirow}
\usepackage{graphicx} 
\usepackage{wrapfig}
\usepackage{hyperref}
\usepackage[super]{nth}
\usepackage[most]{tcolorbox}
\usepackage{siunitx}

\usepackage{array,multirow,graphicx}
\usepackage{nicematrix}
\usepackage{color, colortbl}
\definecolor{aliceblue}{rgb}{0.94, 0.97, 1.0}
\usepackage{float}

\definecolor{light-gray}{gray}{0.95}

\usepackage{xcolor}
\definecolor{my-blue}{HTML}{04d9ff}
\definecolor{my-green}{HTML}{DAF7A6}
\usepackage{ragged2e}
\newcolumntype{L}[1]{>{\RaggedRight\hspace{0pt}}p{#1}}
\newcolumntype{R}[1]{>{\RaggedLeft\hspace{0pt}}p{#1}}

\title{Modeling the Performance of the Burevestnik Nuclear-Powered Cruise Missile }

\author[1,2]{Jake J. Hecla}
\author[1]{R. Scott Kemp}

\affil[1]{Massachusetts Institute of Technology, Department of Nuclear Science and Engineering. Cambridge, Massachusetts, USA}
\affil[2]{Massachusetts Institute of Technology, Department of Aeronautics and Astronautics. Cambridge, Massachusetts USA}

\affil[*]{Corresponding authors: Jake J. Hecla }

\begin{abstract}

In the last decade, Russia's strategic arsenal has pivoted towards a reliance on exotic nuclear-weapon delivery systems. 
One such system, the Burevestnik (NATO: 9M730) is claimed to be a nuclear-powered, nuclear-armed cruise missile capable of nearly indefinite flight. The air-breathing nuclear propulsion system used in this missile is unique, and its attributes are generally unfamiliar to both the aerospace and nuclear-security communities. To better understand the Burevestnik, and the potential of air-breathing nuclear propulsion systems generally, we have developed a nuclear-aircraft modeling toolkit capable of constraining the missile's performance characteristics. Using this framework, we conclude that the Burevestnik is a subsonic cruise missile system measuring $9.5 \pm 0.32$~m in length, with a $5.6 \pm 0.18$~m wingspan, likely powered by a direct-cycle nuclear turbojet (our calculations almost entirely exclude the possibility of a nuclear ramjet). Under these assumptions, our models predict a reactor thermal power of $4.3\pm 1.3$~MWth at cruise, with peak power demand during climb and terminal maneuvering exceeding $15$~MWth, which may be met with a supplemental chemical interburner. Monte Carlo simulations show that escaping neutrons will generate in excess of 5~TBq of gaseous radionuclides per MW-hr of flight, including isotopes such as $^{41}Ar$, $^{85m}Kr$, $^{83m}Kr$ and $^{14}C$, some of which may be detectable using existing monitoring networks.

\end{abstract}
\begin{document}

\maketitle
\thispagestyle{empty}

\section{Introduction}




Nuclear power for the propulsion of aerospace systems is an old concept, first envisioned by Fermi in the mid-1940s. Between the 1950s and the 1970s, the United States and Soviet Union 
carried out testing of airborne non-propulsion reactors and ground testing of nuclear turbojets to evaluate the feasibility of nuclear aircraft. Concepts explored during this era included crewed unlimited-range bombers \cite{layman1962general}, supersonic low-altitude intercontinental cruise missiles \cite{reynolds1962pluto}, unlimited-range submarine hunting aircraft \cite{comassar1962general}, ultra-heavy ground effect vehicles \cite{ADA294979}, and spacecraft propelled by means of nuclear detonations \cite{everett1955method}.
None of these programs resulted in an airborne nuclear-propelled craft,\footnote{Several aircraft have been operated with MW-scale reactors aboard for shielding testing purposes, though they provided no motive power. } and most programs were abandoned by the mid-1970s. 
Publicly-disclosed research on nuclear-powered aircraft was sparse following the initial wave of interest \footnote{Allegations of the use of nuclear propulsion in the Quartz / AARS have been made, though with little firm primary-source backing.}, with occasional paper studies such as the Northrup Grumman--Sandia UP3S program \cite{dron2012unmanned} and the Phantom Works--Sandia \textquote{Extremely Long Endurance Covert UAV} concept \cite{vickers2004revolution}. This ebb in research ended in 2018 when Russia announced the development of a nuclear-powered cruise missile called the Burevestnik. This return to aircraft nuclear propulsion merits a re-examination of the potential capabilities and challenges of aerospace reactors.

Burevestnik is not the first nuclear-powered cruise missile concept. In 1961, the U.S.\ Air Force sought to develop a nuclear ramjet-based supersonic land-attack cruise missile under the moniker \textit{Project Pluto}. Pluto's concept of operation was to fly at treetop level at Mach 3.5 and dispense nuclear weapons along its flight path by performing ``pop up'' maneuvers, striking geographically separate targets hours or days apart. Both a subscale and 460~MW full-scale engine were tested in the Idaho desert, proving the viability of the concept \cite{reynolds1962pluto}. Prior to Pluto, a more conservative effort was undertaken to perform a design study for a direct cycle nuclear propulsion system for the U.S.\ Snark intercontinental cruise missile. In conventionally-powered guise, the Snark was a 21,840~kg turbojet-powered subsonic ICCM which served between 1959--61. It was to be retrofitted with the nuclear A129 engine to provide up to 200~hr of duration at 0.9~Mach. Though differing in scale from the Burevestnik, this system shares the same general concept of operations, engine layout, engine cycle and approximate cruise speed. The A129-Snark study is the only publicly disclosed design of a \textit{subsonic} nuclear-powered cruise missile system prior to Burevestnik. Both the Pluto program and the nuclear Snark projects were abandoned, not for technical reasons, but because the U.S.\ Congress felt nuclear propulsion lacked a well-defined military purpose. 

Today, the bulk of the literature on uncrewed nuclear-powered aircraft is published in China\cite{wen2023numerical,deng2023conceptual,zhou2021investigation,zhao2025numerical,deng2022coupled,lu2023numerical,liao2024precise,bai2019heat,cheng2024novel}suggesting broad-ranging interest in the concept. Of the papers published between 2018 and 2026, all have been computational and there have been no public references to experiments. These papers have explored applications of nuclear propulsion ranging from intelligence gathering aircraft to the propulsion of high-altitude supersonic aircraft.

To better understand Burevestnik, its capabilities, and the prospect for follow-on systems, 
we developed a toolkit that couples drag estimation, engine cycle modeling, and neutronics. The outputs can place constraints on numerous nuclear aircraft performance parameters, providing insight into the engineering challenges involved in developing such systems, as well as the limitations of their flight envelopes and utility for various missions. The model uses openly available data and does not rely on any non-public materials or methodologies.

\section{Background}
\subsection{History of Russian Aircraft Nuclear Propulsion}
Russian interest in air-breathing nuclear propulsion dates to the 1950s \cite{garthoff2016swallow}, and was fueled by the same aircraft range and speed concerns that drove contemporary U.S.\ efforts. The USSR developed a program for nuclear aircraft propulsion under the design bureaus of Myasischev, Kuznetsov, Lyulka and Tupolev beginning in 1955. This effort began with the development of a flying testbed reactor, dubbed the Tu-95 LAL (Letayushchaya Atomnaya Laboratoriya or \textquote{flying atomic laboratory}) to evaluate shielding \cite{SovietSecretBombers}. Designs for a Tu-95-based nuclear-powered bomber, dubbed the Tu-119, were made using the shielding measurements performed on the Tu-95 LAL platform. However, the Tu-119 did not progress beyond the design stage. Disclosures to the IAEA indicate the existence of two other experimental aircraft-propulsion reactors, the VVRL-02 and VVRL-03, constructed by NIIP Lytkarino and apparently patterned on a closed-cycle pressurized-water design \cite{podvig2017use}. VVRL-02 achieved first criticality in 1974. Little else is known about these platforms, except for the fact they used HEU, and were decommissioned between 2003 and 2011. IAEA declarations state both are \textquote{transportable} pressurized-water type reactors with an output of 100~kWth.\footnote{There exist published claims of programs seeking to develop submarine-hunting aircraft using nuclear turboprops well into the 1970s and 1980s. An AN-22, BuNo CCCP-08838 was allegedly equipped with a neutron source, and later a 3-kW research reactor as part of a program to study the possibility of extending the range and power budget of ASW aircraft. However, these claims are thinly supported: despite several articles on defense aviation websites declaring the existence of this project, none cite primary sources \cite{aviatsiya_vremya_1997_5_article}.}

The Burevestnik\footnote{Burevestnik is a type of a bird, known as a Storm Petrel in English. The Russian name translates literally to \textquote{storm messenger.} Burevestnik has a symbolic meaning from a famous poem by Maxim Gorky, \textit{\textquote{The Song of the Storm Petrel}} (\textit{Песня о Буревестнике}, 1901). In the poem, the burevestnik is portrayed as a fearless creature willing to fly when all other birds hide. It became a symbol of the Russian Revolution indicating fearlessness in the face of revolutionary upheaval.} is a ground-based, nuclear-powered, nuclear-armed strategic cruise missile system unveiled in March 2018, followed by a series of propaganda videos\cite{Hruby_RussiaNewNuclear_2019, Minoborony_RaketaYaEU_2018, Minoborony_Burevestnik_2018}. It is an intercontinental cruise missile (ICCM), a concept not implemented since the U.S.\ Snark missile of the late 1950s and with no precedent in the Russian armed forces. It is a product of the All-Russian Scientific Research Institute of Experimental Physics (VNIIEF), the organization primarily responsible for Russian nuclear weapon design. Russian media reports claim that the development program dates to 2001 \cite{Izvestia_Burevestnik_2019}, and that the powerplant can trace its technical lineage to the work of M. Lyulka \cite{Leonkov_Burevestnik_2019}. The Burevestnik is powered by a compact nuclear propulsion system designed by the KB-12 group (special topics division) within VNIIEF \cite{KANAL16}. In contrast to chemical-fuel cruise missile systems, Burevestnik nuclear plant gives it the ability to fly for days to weeks. In theory, Burevestnik could loiter, map air defenses, and attack via unexpected routes, using terrain to escape radar detection and evade missile defenses. Informal statements by military proxies seem to indicate that the system may use some level of autonomy to update trajectory information in response to countermeasures. Generally, it is considered that the system is intended to be land-based, though there has not been an official confirmation of this design choice.


The Burevestnik has been characterized in Russian publications as a response to evolving U.S.\ missile defenses. Informal statements by Russian military proxies have characterized it as a \textquote{revenge} weapon\cite{Leonkov_Burevestnik_2019}, suggesting a second-strike role. However, it should be noted that the system's ability to loiter in radar clutter and strike at a precisely predetermined time does not rule out the possibility of a first-strike role. Though no clear statements about the intended warhead, though Russian military bloggers and former officials claim it would be a multiple-megaton thermonuclear weapon \cite{Novikov_Desyatki_2023}. 

Reports from the Nuclear Threat Initiative have stated that more than a dozen failed launches were carried out prior to 2020, with at least two described as partial successes by the US intelligence community \cite{Hruby_RussiaNewNuclear_2019}.  In 2019, an accident at the Nyonoksa military test site brought the system to international attention. In the early hours of August 8th, an explosion took place during the recovery of a Burevestnik test article, killing at least five Rosatom scientists and generating a seismic signal detected in Norway \cite{krzyzaniak2019nenoksa}. In the hours following the explosion, Roshydromet (the Russian Federal Service for Hydrometeorology and Environmental Monitoring) released a statement which was later deleted indicating the detection of short-lived fission products in the resulting plume in Severodvinsk \cite{Peshkov_Zagryaznenie_2019}. Norwegian radioisotope measurements were suggestive but not decisive, indicating an airborne $^{131}$I spike in the days following the accident, but without additional fission product traces. Russian CTBTO-IMS detectors were strategically deactivated along the anticipated plume path over the following week\cite{Bellona_Detectors_2019}. Doctors in contact with the injured Rosatom employees reported to the media that they had been contaminated with $^{137}Cs$ \cite{gershkovich2019exclusive}, again indicative of a reactor accident. The presence of fission products in the radioisotope data released by Roshydromet, combined with U.S.\ intelligence community statements \cite{Macias_IntelRussianExplosion_2019} strongly supports the assertion that the system was being tested with a reactor aboard, and that a criticality accident took place during recovery resulting in the fatalities. This may not have been the first failure of the Burevestnik: some have linked the release of Ru-106 in the Urals in 2017 to the testing of the system \cite{Hruby_RussiaNewNuclear_2019}.

In October 2023, The Russian Ministry of Defense (MoD) claimed that they had carried out successful flights under nuclear power  \cite{Plummer_PutinMakes_2023,Kremlin_Posechenie_2025,Eckel_PutinTestsTorpedo_2025}, though without presenting further footage or technical specifications. Two years later, they again claimed a successful Burevestnik flight test, this time specifying that the flight spanned 15~h, and covered a range of 14,000~km \cite{Kremlin_Posechenie_2025,Eckel_PutinTestsTorpedo_2025,Lebedeva_Burevestnik_2025}. Norwegian national security officials later confirmed that a test took place in Novaya Zemlya in that time window, and that the system flew \textquote{significantly longer than before,}\cite{Eckel_PutinTestsTorpedo_2025} though without citing exact range or duration estimates. These claimed test dates also correspond with public ADSB data showing movement of U.S. WC-135 Constant Phoenix aircraft to the region.

\subsection{Fundamentals of Nuclear Propulsion}

Air-breathing nuclear propulsion systems generate thrust using the atmosphere as the primary reaction mass, and a nuclear reactor as the source of enthalpy. These systems are distinct from other nuclear-thermal propulsion systems such as nuclear thermal rockets in that they do not rely on internal stores of propellant. Air-breathing nuclear engine designs include turbojets, turbofans, turboprops, ramjets, scramjets and pulse-jets. To date, the only air-breathing nuclear engines that have been built and tested are nuclear turbojets and ramjets.

Turbine engines generate power using the Brayton cycle. The idealized Brayton cycle entails the isentropic compression of a working gas, isobaric heat addition, isentropic expansion, followed by isobaric heat rejection. Work extracted in the expansion step is used to drive the compression step and/or an external load. In an open-cycle turbojet or turbofan engine, the working fluid is atmospheric air which is expelled to create thrust. While the heat addition step is conventionally performed by combusting a hydrocarbon, the heat can be added by any exothermic process including a nuclear reactor. Modern turbomachines typically operate at overall pressure ratios (ratio of the pressure at the compressor outlet divided by the pressure at the compressor inlet, termed \textquote{OPR}) in the range of 15--60. At the turbine inlet, gas temperatures in such systems can range between 1200--1700~K. For smaller turbojet and turbofan engines such as those used in cruise missiles, OPRs tend to be under 20, and turbine inlet temperatures under 1600~K. 


For nuclear propulsion, the means of heat transfer to the reaction mass (air) can be classifed as either \textquote{direct cycle} or \textquote{indirect cycle.} Direct-cycle systems use atmospheric air as the primary coolant for the reactor, passing it through the core and heating it by direct contact with fuel elements. This approach offers lower total powerplant mass (in most size regimes), but irradiates the air, generating an irreducible radiological signature from neutron capture on atmospheric gases and aerosols. In direct-cycle systems, fission products may diffuse into the gas stream, and cladding materials can erode, resulting in the emission of further radioactive byproducts. Likewise, structural materials subject to neutron flux (for example, compressor or turbine blades) may become activated and shed radioactive wear particles. Direct-cycle reactors additionally may be vulnerable to neutronic disturbances introduced by moisture, air-density changes and foreign debris in the intake air stream. The U.S.\ Aircraft Nuclear Propulsion program tested both turbojets (HTRE series) and ramjets (TORY series) based on the direct-cycle layout, with mixed success in maintaining exhaust-gas cleanliness\cite{friesen1995radiological}.

Indirect-cycle systems use a circulating intermediate heat-transfer medium such as a molten metal, molten hydroxide, molten salt, or pressurized gas to transmit heat to the turbomachinery. No direct contact exists between fuel and propellant in such designs, and the resulting isolation of the core offers improved fission product retention. This, in turn, allows for higher maximum fuel temperatures and potentially higher power densities. However, the added mass of the coolant, pumping system, pressure-compensation system, heat exchangers, and pressure vessel typically results in considerably more overall mass, a disadvantage in comparison with direct-cycle systems. A wide diversity of indirect-cycle layouts are possible, including gas-turbine driven ducted fans\cite{thompson1979compact}, Rankine-cycle systems, and hybrid-electric layouts\cite{cheng2024novel}. Early indirect-cycle work focused primarily on systems with solid fuel cooled by molten metals (typically Na, K, and Li) and liquid-fuel reactors using molten salts. Design work performed by NASA has suggested that megawatt-scale, metal-cooled systems are possible with reactor and pumping equipment masses around 1000~kg \cite{kelly1964snap}. In recent years, gas-cooled systems have been examined by both U.S.\cite{dron2012unmanned} and Chinese groups \cite{liao2024precise} for UAV propulsion.


\begin{figure}[htbp!]
    \begin{center}
    \includegraphics[width=.96\columnwidth]{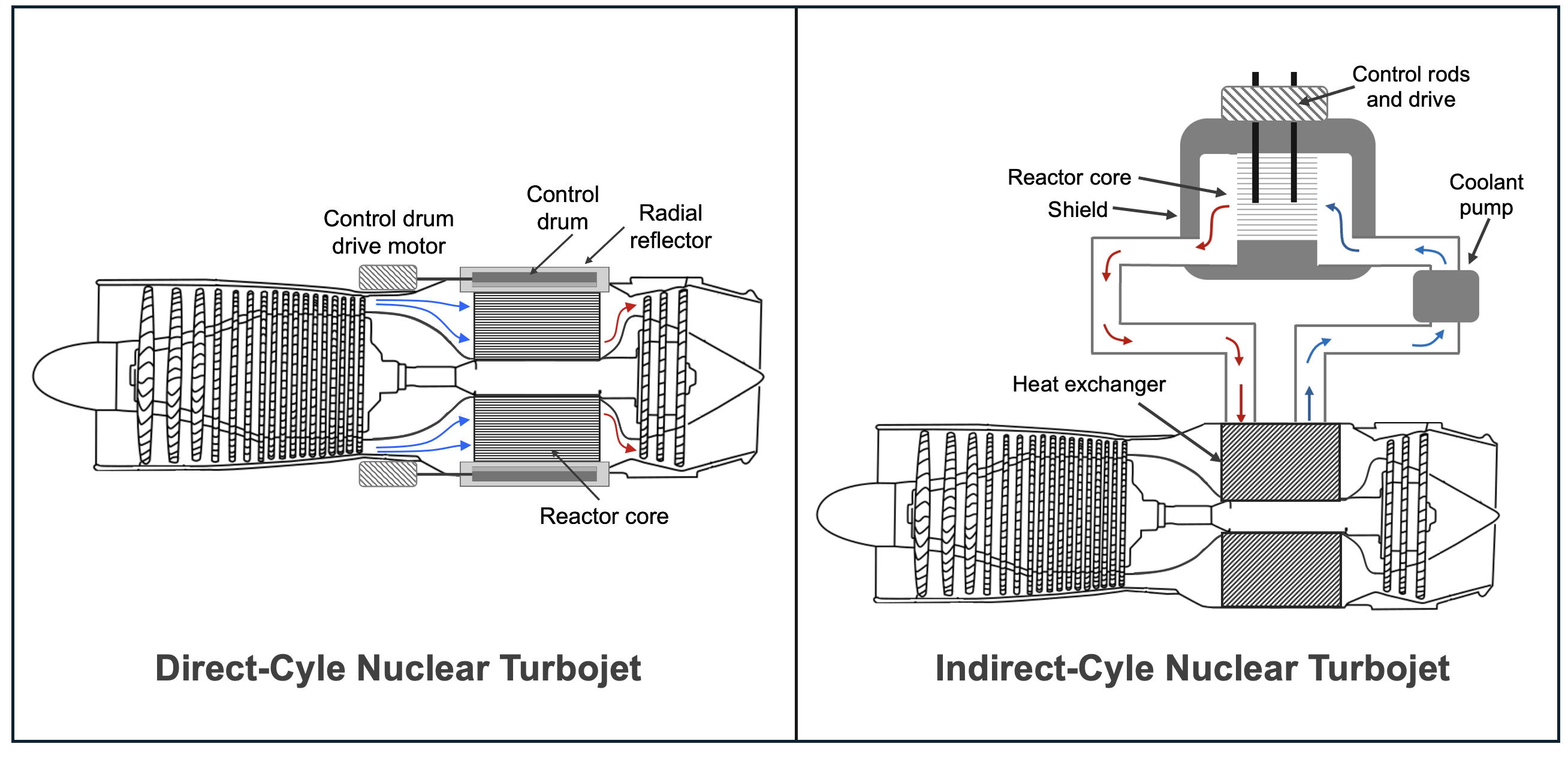 }
    \end{center}
    \vspace*{-5mm} 
    \caption{\textbf{Left:} A direct cycle nuclear propulsion system replaces the combustor in a conventional turbomachine with a reactor, and offers the simplest means of creating a nuclear propulsion system. \textbf{Right:} An indirect cycle system replaces the combustor with a heat exchanger driven by a reactor. This design has improved isolation of fuel from the propellant, and therefore has a likely improved safety profile, as well as less exhaust radioactivity, though at the cost of complexity and mass.}
    \label{fig:IndirectCycle}
\end{figure}

Most air-breathing nuclear propulsion concepts retain significant commonality with hydrocarbon-based turbomachines used in aircraft propulsion and power generation. However, there are several notable differences. In contrast to combustion-powered systems, nuclear systems tend to see high pressure losses in the heat exchanger since heat must be transferred across solid-gas interfaces. Additionally, the materials of the heat exchanger (or reactor fuel) generally tolerate lower temperatures than combustion systems, which lowers achievable turbine-inlet temperatures. Finally, reactors change power levels more slowly, which leads to slow transient response; to alleviate this issues, many air-breathing nuclear propulsion concepts have used chemical-fuel \textquote{interburners} positioned behind the reactor to allow for rapid power additions.

\section{Modeling the Burevestnik}
\subsection{Flight Parameter Estimation and Engine Sizing}
To better understand the Burevestnik missile system, we use openly available data coupled to a modeling pipeline to estimate its size, mass, thrust, engine layout, and likely radiological signatures. We assume that the publicity videos of the Burevestnik faithfully represent the size and shape of the missile system, and that the system derives most of its motive power from nuclear energy. The body of openly-available information on the Burevestnik consists of a small handful of Russian Ministry of Defense (MoD) official statements, four MoD video clips showing missile hardware, amateur photographs of recovery equipment following the August 2019 criticality accident, and a handful of satellite images taken of the Pankovo and Nyonoksa test sites during missile testing campaigns \footnote{To dimension and model the missile within this limited information environment, we draw data from the following sources: \href{https://web.archive.org/web/20191029155918/https://www.svoboda.org/a/30144456.html}{amateur photographs} of accident debris republished by Radio Svoboda, \href{https://web.archive.org/web/20180323082045/https://www.youtube.com/watch?v=Xr7alYwCznQ}{Russian MoD video no. 1} containing: \textquote{close-up launch} footage (0-2 s), \textquote{distant launch} footage (4-12 s), and \textquote{flight} footage (12-14 s), \href{https://web.archive.org/web/20190202015837/https://www.youtube.com/watch?v=okS76WHh6FI}{Russian MoD video no. 2} containing \textquote{factory} footage (12-86 s), and Planet Labs and Airbus DS overhead imagery provided by both the \href{https://planesandstuff.wordpress.com/2022/09/17/burevestnik-actually-could-be-ready-to-test/}{James Martin Center for Nonproliferation Studies} and \href{https://www.armscontrolwonk.com/archive/1210186/burevestnik-testing-to-resume/}{in blog posts by Tony Roper}.}. 

\subsubsection{Missile Dimensions from Fiducial Objects}

\begin{figure}[htbp!]
    \begin{center}
    \includegraphics[width=.940\columnwidth]{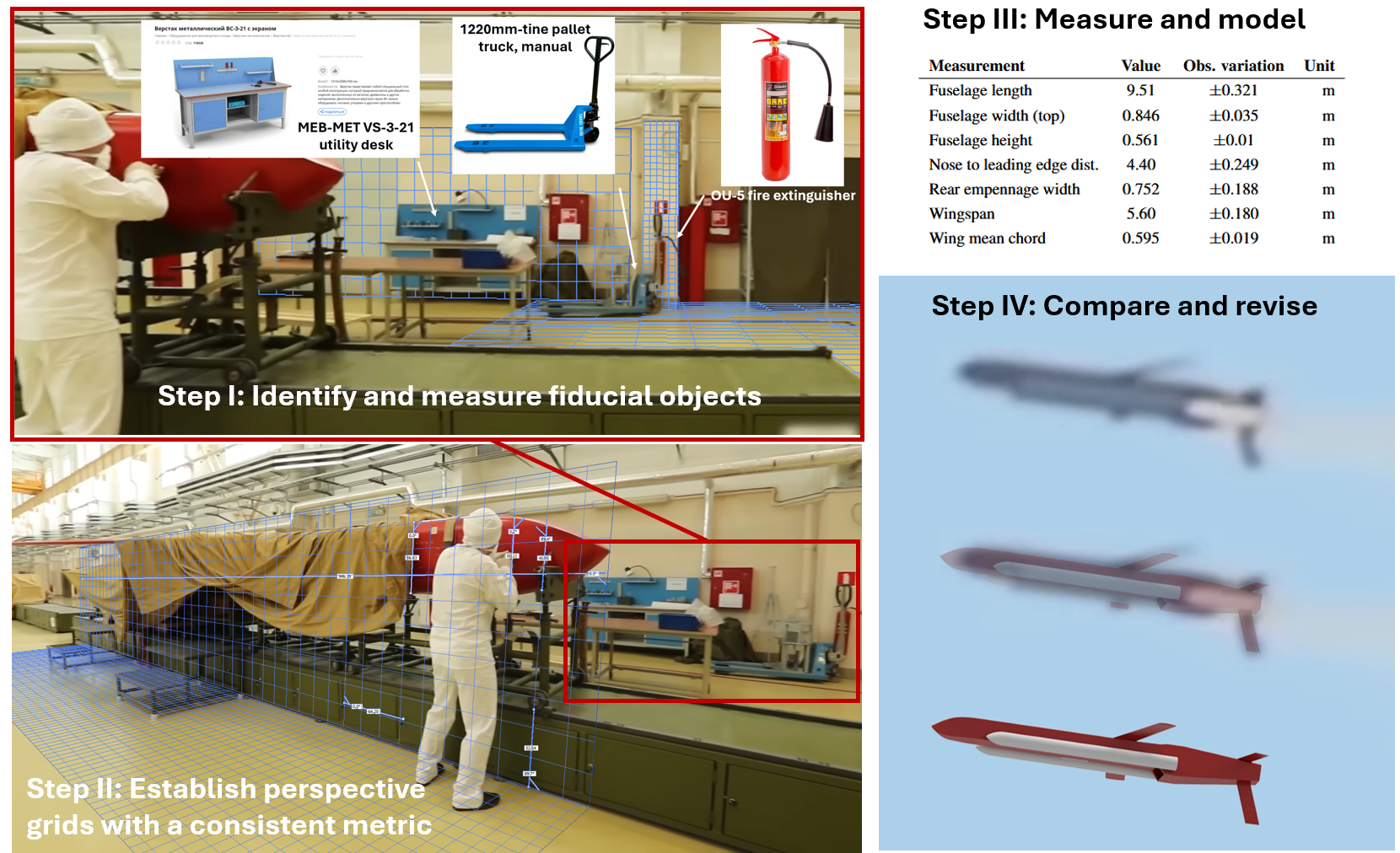 }
    \end{center}
    \vspace*{-2mm} 
    \caption{This series of figures illustrates the process by which fiducial objects, perspective grids, and modeling software have been used to construct dimensionally-accurate 3D representations of the Burevestnik. \textbf{Top-left:} Fiducial objects identified in the “factory” video include the \href{https://web.archive.org/save/meb-met.ru/verstak-metallicheskiy-vs-3-21-s-ekranom.html}{MEB-MET VS-3-21 utility desk}, a 1220~mm manual pallet jack (comparable to the \href{https://web.archive.org/web/20250802211246/https://www.bluegiant.com/products/ergonomics/industrial-trucks/manual-pallet-trucks/fpt-55/}{Blue Giant FPT-55}), and an \href{https://web.archive.org/web/20230606130617/https://pozhtechnika.com/production/co2-carbon-dioxide-fire-extinguishers-inei/co2-carbon-dioxide-fire-extinguisher-inei-5kg-bce-4489/}{OU-5 multipurpose fire extinguisher}. \textbf{Bottom-left:} The perspective grid containing the fiducial objects has been extended to include the missile support structures. \textbf{Top-right:} Our extracted geometry measurements are shown with uncertainties derived from the distribution observed when repeatedly measuring the selected object using differing video frames and perspective grids. \textbf{Bottom-right:} A draft CAD model derived from the factory measurements is shown superimposed on the flight footage to check consistency. }
    \label{fig:Measurement}
\end{figure}

The first step in our analysis was to build a dimensionally accurate model of the Burevestnik from videos and photographs using photogrammetry\cite{cia-photographic-measurement}. These measurements were based on fiducial objects, among them work benches, standardized fire extinguishers, and a manual pallet truck. We identified the manufacturer of each object and obtained dimensions from which we established perspective grids. These grids were then used to measure the missile and its surroundings, as illustrated in Figure~\ref{fig:Measurement}. The bulk of the airframe dimensions were extracted from the \textquote{factory tour} video, as it provides multiple views of Burevestnik test prototypes in a consistent setting. Not all relevant portions of the geometry are visible in the video as the result of drop-cloths which obscure certain parts of the airframe. However, these unknown dimensions can be constrained using low-resolution imagery in the \textquote{flight} and \textquote{distant launch} clips, combined with reasonable assumptions about geometry. For example, transportation canisters shown in the factory provide one such constraint: the width of the aft fuselage cannot be wider than the missile canister width, nor can the body height plus the prominence of the fixed intake exceed the interior height of the canister. Our estimates of the vehicle's dimensions are shown in Table~\ref{tab:measurements}.

From the in-flight footage, we conclude that the missile uses a bottom-mounted horizontal stabilizer, a bottom-mounted air intake\footnote{There is some disagreement as to whether the intake is on the top or bottom of the fuselage. There is an intake-like cavity visible in some overhead shots in the factory video. However, the in-flight footage shows a bottom-mounted intake. We interpret the cavity on the top as the port through which the reactor is installed. As the factory shown is not suitable for nuclear operations, it must happen at a later assembly step.} and a top-mounted horizontal stabilizer. Further, in-flight footage allows redundant estimation of wing shape, wing-root location and wing sweep to bolster our prior \textquote{factory} measurements. While the flight and launch videos are subject to blur and washout from movement and booster flair, and therefore are less precise than the factory footage, they allow us to build a complete and consistent picture of the dimensions of the system from multiple views. A subset of the extracted system measurements are given in \ref{tab:measurements}.

The overall planform we reconstruct from the video footage strongly suggests that Burevestnik is a subsonic system. It has long, relatively high aspect ratio wings and aft control surfaces with moderate sweep angles. The fuselage is blunt, and has a relatively constant cross-sectional area as shown in Table~\ref{tab:measurements}. These features are consistent with subsonic flight, and a traditional cruise missile \textquote{nap of the earth} trajectory. The planform we reconstruct strongly resembles existing subsonic systems such as the Kh-69, which shares a blunt squared-off cross section, and the Kh-101, which has a similar empennage and wing shape. By contrast, an efficient supersonic design would have a thin, highly-swept, low aspect ratio wing, and an area-ruled fuselage. We also tracked booster-exhaust disturbances in video footage, which suggest that the prototype missile was flying around 0.79~Mach (see the appendix for further details).

\begin{figure}[htbp!]
    \begin{center}
    \includegraphics[width=.80\columnwidth]{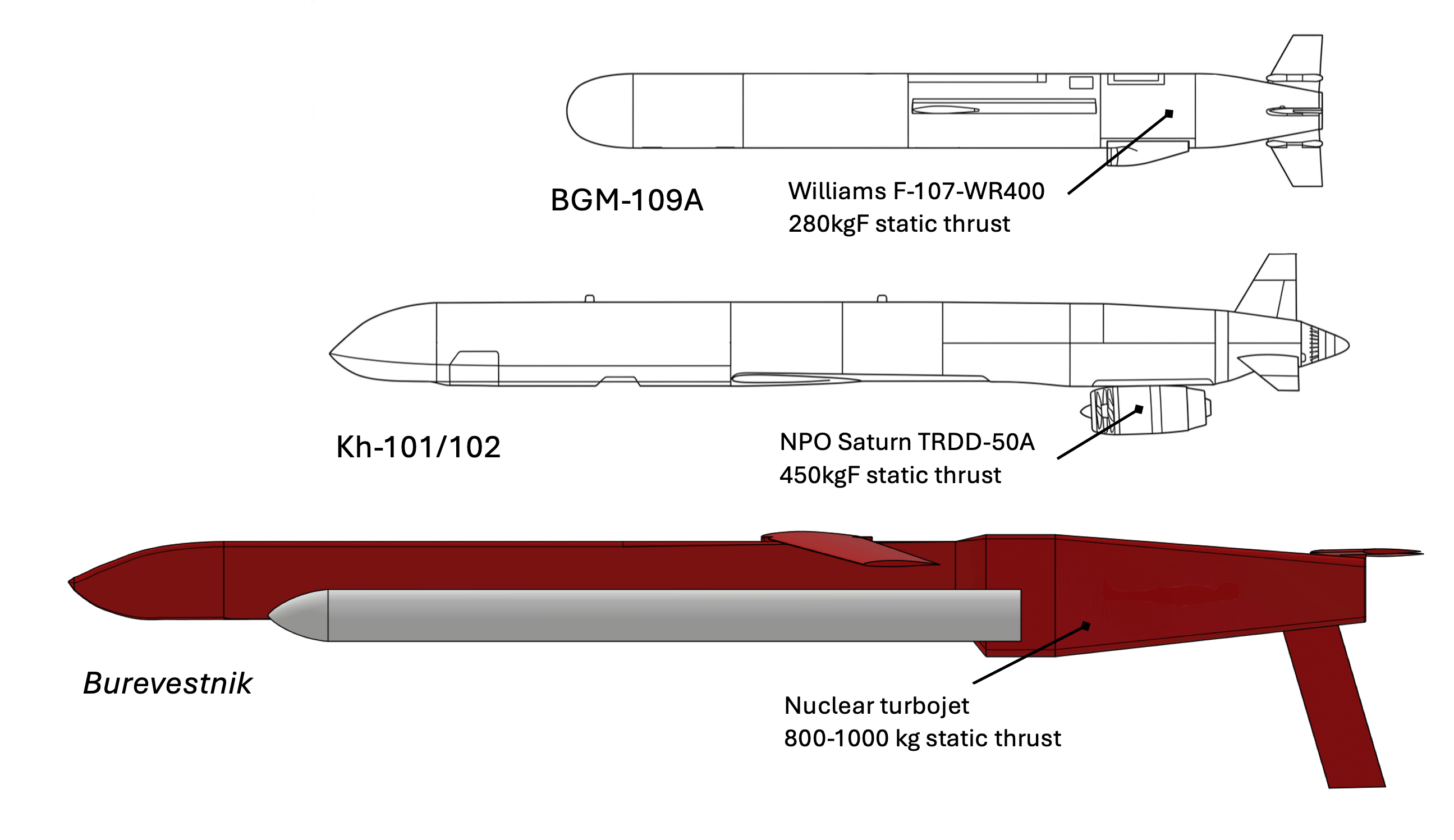}
    \end{center}
    \vspace*{-10mm} 
    \caption{The model of the Burevestnik extracted from video evidence measures approximately 9.5~m in overall length, with a wingspan of approximately 5.6~m. SRBs have been included here for clarity, though they are not included in the following drag-workup procedures. Note that this diagram (and the following descriptions) assume a bottom-mounted intake, which matches the available evidence best.} 
    \label{fig:MissileComp}
\end{figure}

\begin{figure}[htbp!]
    \begin{center}
    \includegraphics[width=.840\columnwidth]{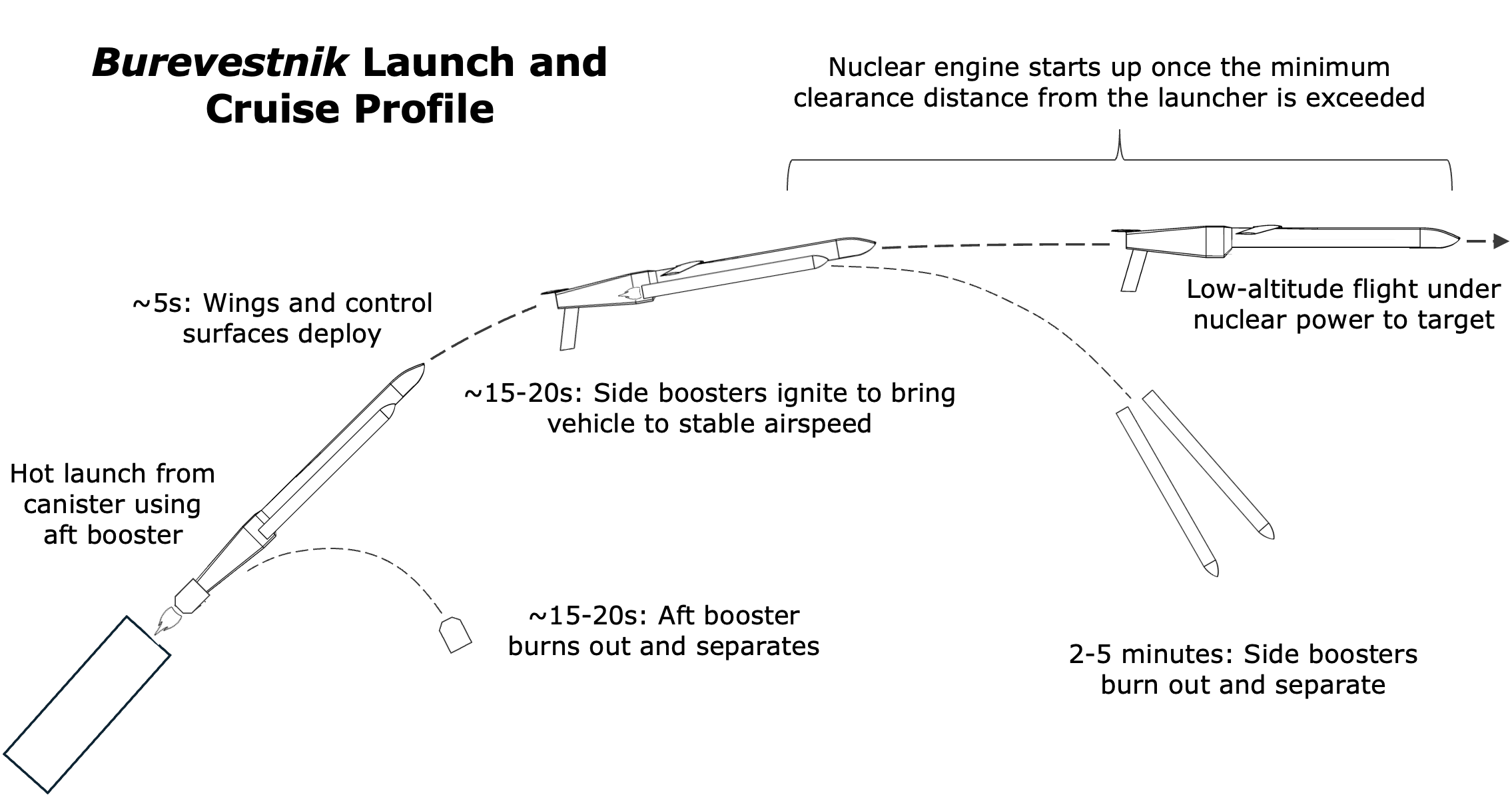}
    \end{center}
    \vspace*{-2mm} 
    \caption{The notional Burevestnik concept of operations consists of launch using a kicker, then transitioning to solid rocket booster power. This then allows a slow spool-up to nuclear cruise at high-subsonic speeds. Alternatively, the SRBs may be for testing purposes only, and the nuclear engine system may instead use hydrocarbon fuels to slowly taper from conventional power to nuclear power.}
    \label{fig:ConOps}
\end{figure}

Based on the launch video evidence, we judge that the Burevestnik shares a relatively conventional start sequence. Frame-by-frame review of both launch clips show that the missile leaves its canister propelled by a single aft-mounted solid rocket \textquote{kicker.} Most ground-launched cruise missiles use a similar hot-launch arrangement to provide enough time and airspeed for their turbofan engines to spool up and begin providing thrust. In the Burevestnik in-flight footage, the kicker is now absent, and twin solid rocket boosters (SRBs) are shown actively burning. These may be intended to get the missile to a safe distance from the launch complex before the nuclear propulsion system reaches criticality, or to allow time for a gradual engine spool-up. Alternatively, the SRBs shown in the video footage may be stand-in propulsion for aerodynamic testing without a nuclear powerplant. This sequence is illustrated in Figure~\ref{fig:ConOps}.


\FloatBarrier
\subsubsection{Drag Modeling}

To estimate of the aerodynamic characteristics of the Burevestnik 
we use the \textquote{Delta Method,} a semi-empirical drag buildup technique developed by Lockheed and NASA that is valid in the high sub-sonic regime \cite{feagin1978delta}. The lift and drag are defined as: \begin{equation}
L = \frac {\rho V^2 C_L S_{ref}} {2}  \quad \textrm{and} \quad D = \frac {\rho V^2 C_D S_{ref}} {2},
\end{equation}
In which $\rho$ is the air density, $V$ the airspeed, $S_{ref}$ the reference area, and $C_L$ and $C_D$ the coefficients of lift and drag, respectively. The Delta method estimates the drag coefficient by aggregation of individual contributions, and is a function of the lift coefficient:    \newline
\begin{equation}
C_D = \underbrace{C_{D_F} + \Delta C_{D_C}+\Delta C_{D_F} + \Delta C_{D}}_{\textrm{drag at zero lift}}+\underbrace{\Delta C_{D_P} + \frac{C_L^2}{\pi AR}}_{\textrm{drag due to lift}} 
\end{equation} 
$C_{D_F}$ is the drag due to viscous shear stresses of turbulent flow across flat-plate approximations for the body, wing and tail; $\Delta C_{D_F}$ is a set of corrections to account for real-world skin friction arising from factors such as surface roughness; $\Delta C_{D_C}$ is the sum of compressibility drag on the wing and fuselage (with interference terms); and $\Delta C_{D}$ miscellaneous drag from elements such as antennae, which is neglected for our purposes. For the lift-based components, $\Delta C_{D_P}$ is wing pressure drag, which depends on both $\Delta M$ (deviation from design Mach) and $\Delta C_L$ (deviation from design lift coefficient). Throughout,  $AR$ represents wing aspect ratio, with an Oswald efficiency factor assumed to be unity (as required for this method). Each drag buildup term can be extracted graphically from semi-empirical relations given in Feagin et al.\cite{feagin1978delta} that relate physical parameters to each drag term. Since the Delta Method method was designed for manned aircraft, we validated it for cruise-missile-like planforms by applying it to Boeing's Long Range Conventional Strike Weapon (LRCSW) and to the AGM-109A Tomahawk Land Attack Missile (TLAM). Agreement with empirical flight data were good. Details are given in Appendix~\ref{app:delta-verify}. 

\begin{figure}[htbp!]
    \begin{center}
    \includegraphics[width=.90\columnwidth]{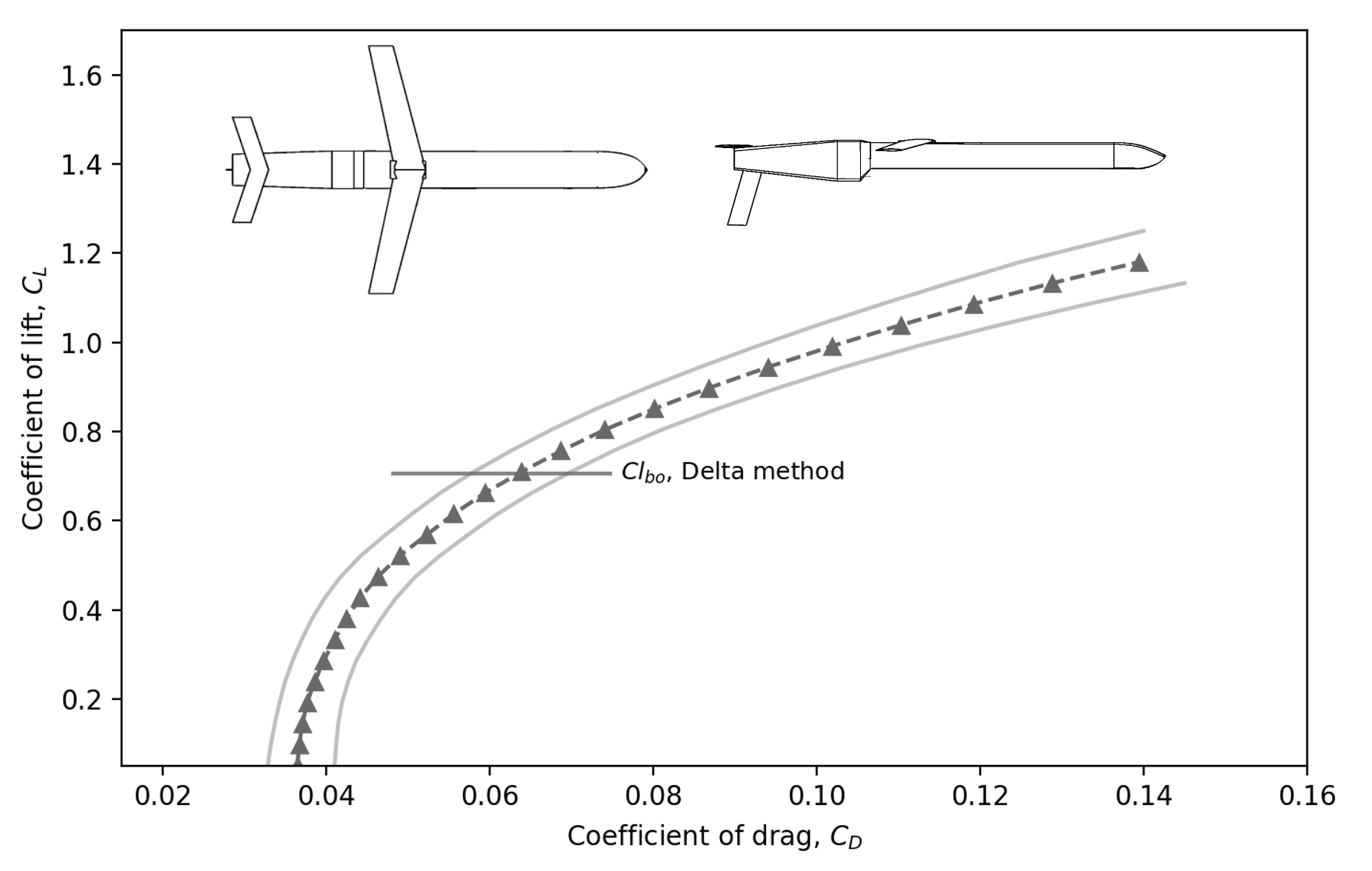 }
    \end{center}
    \vspace*{-2mm} 
    \caption{Drag Polar estimate for the Burevestnik (with propagated uncertainties) at 0.85 Mach. The $C_L/C_D$ properties are slightly better than the Tomahawk cruise missile, but somewhat less than proposed advanced cruise missile designs \cite{feifel1992propulsion}. $Cl_{bo}$ is the predicted lift coefficient at buffet onset. Our benchmark suggests the Delta Method may under-predict $C_L/C_D$ beyond this point (see Appendix~\ref{app:delta-verify}).} 
    \label{fig:DragPolar}
\end{figure}

Application of the Delta Method to the Burevestnik requires assumptions to be made about the flight regime and certain aerodynamic parameters. These assumptions introduce error that are not easily quantified. We have assumed that (1) the Burevestnik uses a supercritical airfoil, (2) the horizontal stabilizers have a chord equivalent to the vertical stabilizer, and (3) the low-mounted engine intake has been skillfully designed with a pressure recovery factor near unity. These factors have been listed in order of decreasing impact on the drag polar. In the drag polar result presented in Figure~\ref{fig:DragPolar}, a 3\% methodological uncertainty has been added to $C_D$ to represent the typical deviation between flight-test data from subsonic aircraft and drag values predicted by the Delta Method. Uncertainty from the photogrammetry measurements (Table~\ref{tab:measurements}) have been propagated and results are presented in Table~\ref{tab:delta-results} alongside a calculation of the minimum drag.

\begin{table}[!htb]
    \begin{minipage}[t]{0.45\linewidth}
      \caption{Table of measurements of Burevestnik developed using the approach described in \ref{fig:Measurement}. \newline Note: we assume that the sweep angle of the \newline horizontal stabilizer (marked with *) is equal to  \newline that of the vertical stabilizer. }
      \label{tab:measurements}
      \centering
        \begin{tabular}{ll}
        \textbf{Component}&
        \textbf{Measurement}
        \\
        \hline
        
        Fuselage length	&	9.51 \pm	0.32~m \\
        Fuselage width 	&	0.846 \pm	0.04~m	\\
        Fuselage height	&	0.561	\pm	0.01~m	\\
        Wingspan	&	5.60	\pm	0.18~m\\
        Wing mean chord &	0.595		\pm	0.02~m \\
        Sweep angle	&	14.8 \pm	$3.8^\circ$ \\
        Wing T/C	&	6.46  \pm $.015~\%$  \\
        Vert. stab. length	&	1.18	\pm	0.16~m  \\
        Vert. stab mean chord	&	0.393	\pm 0.16~m \\
        Vert. stab sweep	&	18.6	\pm $3.5^\circ$ \\
        Horz. stab span	&	2.38	\pm 0.24~m \\
        Horz. stab mean chord	&	0.393*			\pm 0.12~m* \\
        
        \end{tabular}
    \end{minipage}%
    \begin{minipage}[t]{.45\linewidth}
      \centering
        \caption{Table of the Delta Method drag buildup terms calculated from the measurements in the table at left. Values shown have been calculated near design Mach. \newline \newline}
        \label{tab:delta-results}
        \begin{tabular}{ll}
        \textbf{Drag term}&
        \textbf{Value}
        \\
        \hline
        $C_{D_F-wing}$	& {7.67E-03}	\pm {1.29E-04} \\
        $C_{D_F-fus}$	& {1.55E-02}	\pm {5.85E-04} \\
        $C_{D_F-tail}$	& {2.74E-03} \pm	{1.71E-05} \\
        $\Delta C_{D_C-wing}$ &	{1.98E-03}	\pm {5.32E-04} \\
        $\Delta C_{D_C-{int}}$ &  ----------- \\
        $\Delta C_{D_C-fus}$ & {1.20E-03}	\pm {7.37E-04} \\
        $\Delta C_{D}$ & ------------ \\
        $\Delta C_{D_P-wing}$ & {3.50E-03}	\pm {1.63E-03} \\
        \hline	
        $C_{D_{min}} = $	& {0.03639} \quad \pm \quad {0.0038}
        \end{tabular}
    \end{minipage} 
\end{table}




\FloatBarrier
\subsection{Powerplant Requirements}
\label{sec:requirements}

\begin{figure}[htpb!]
    \begin{center}
    \includegraphics[width=.90\columnwidth]{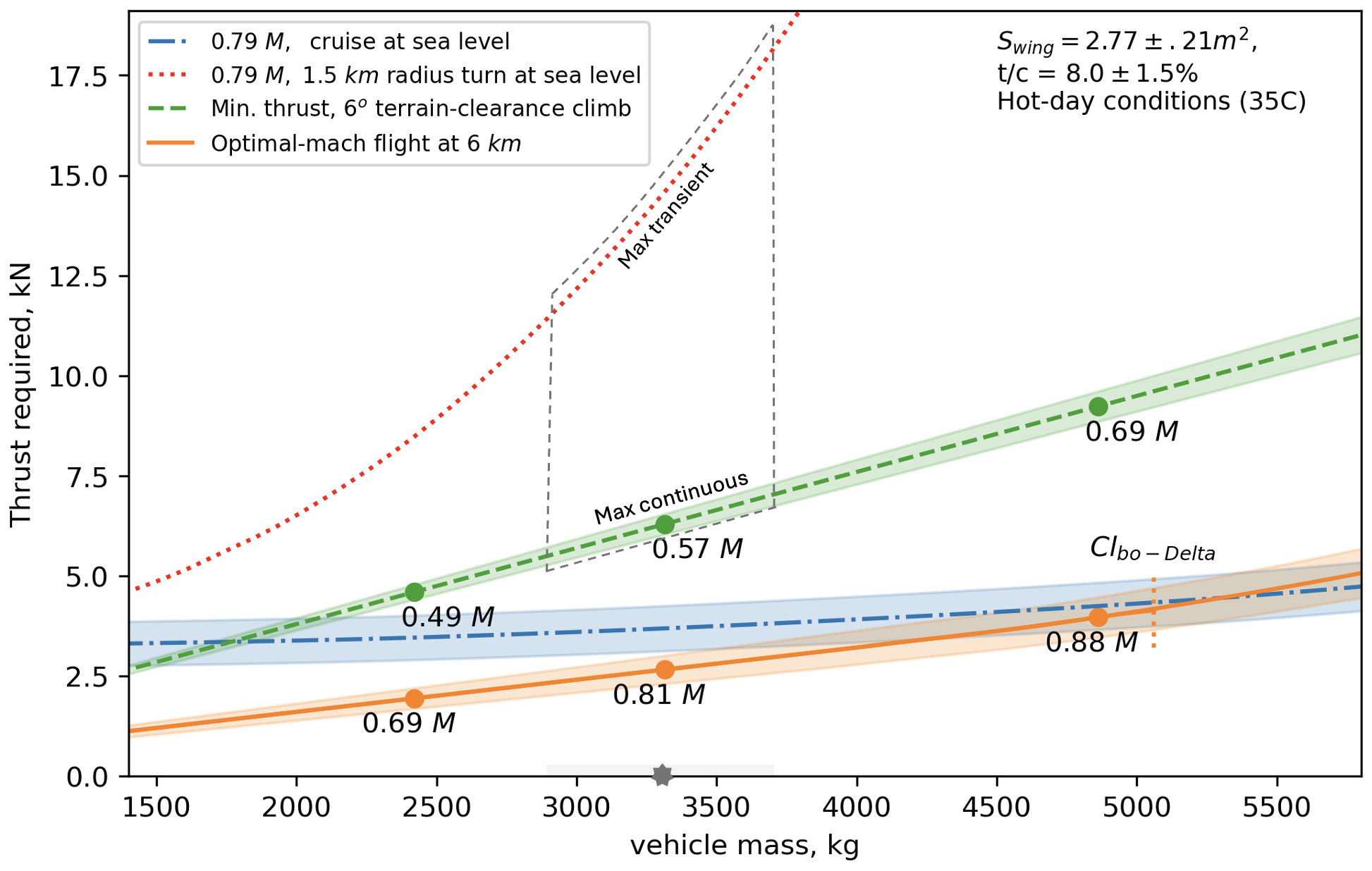 }
    \end{center}
    \vspace*{-8mm} 
    \caption{Thrust requirements for the 3 limiting maneuvers as a function of vehicle mass. The dashed area highlighted in the plot represents the most likely range of mass and maximum thrust values for the Burevestnik resulting from this analysis ($\pm 1~\sigma $ region), with the median denoted by a star on on the x-axis. Note that the propulsion requirements for the 1.5~km-radius turn could be met using a chemical interburner or afterburner, and is thus not a strict requirement of the nuclear system. The nuclear system must \textit{at minimum} provide for cruise and sustained climb. The error in the drag polar driven by design uncertainties has been propagated through in this analysis. As the drag polar uncertainties grow rapidly at high $C_L$ values, we do not quantify them in the 1.5~km-radius turn curve. A reasonable range for $Cl_{max}$ is likely between 0.9--1.3. We have also indicated the predicted buffet onset for high-altitude flight ($Cl_{bo-Delta}$), based on the Delta Method.}
    \label{fig:PropulsionRequirements}
\end{figure}

Given the planform, we judge the flight envelope of the Burevestnik is likely similar to that of conventionally powered cruise missiles, differing only in range/endurance. Under this assumption, we can draw on the existing cruise missile design literature to establish approximate thrust requirements. In \textit{A Conceptual Design Study for a Joint-Service Cruise Missile}, Feifel and Kerkam \cite{feifel1992propulsion}  identify the limiting maneuvers that drive wing and powerplant sizing in heavy, long-range cruise missiles to be:
\begin{enumerate}
    \item Climbing at a 6$^\circ$ angle at optimal Mach, at sea level, in hot-day conditions.
    \item Maintaining level flight at 6,000~m in hot-day conditions to clear high terrain.
    \item Performing a 1,500~m radius turn during terminal phase at sea level in hot-day conditions. 
\end{enumerate}
The thrust needed to perform each of the above maneuvers is calculated using the drag polar as a function of the system mass, following Marchman \cite{marchman2004aerodynamics}. Results are given in Figure~\ref{fig:PropulsionRequirements}. Note that the system mass is a free parameter. 

\subsection{System Mass}
We estimate a plausible system mass based on an empirical wing-load relationship that we derived by regression analysis of 13 missiles with similar planforms and wing-aspect ratios using a methodology similar to Ekker \cite{ekker1994missile}. This analysis yields a best-fit system mass based of $3340 \pm 804~kg$, in which the uncertainty includes both wing-area estimation error and regression-related error.\footnote{Alternatively, assuming wing loading similar to the Kh-101 which draws on shared design personnel and serves a similar strategic role, we calculate a system mass of $3194 \pm 242~kg$.} As before, we benchmarked this method using data from the BGM-109M program, showing robust agreement.

From the system mass budget, one must accommodate the airframe, powerplant, and shielding. The last component requires an estimate of component radiation tolerances, which in turn requires a fairly detailed design. A \textit{very approximate} upper bound on the nuclear-propulsion system mass can be made by arguing from similarity with conventional systems. Literature on hydrocarbon-powered long-range cruise missiles \cite{ekker1994missile} suggests that the upper limit on the fuel + powerplant mass is approximately 40\% of the total system mass. Under the assumption that no chemical fuel is needed aboard Burevestnik, but shielding is; this yields a propulsion-system with shielding mass of approximately 1340~kg. Subtracting 250~kg for turbomachinery and auxiliary systems (from an informal survey of similarly sized engines) leaves approximately 1090~kg (32\%) for the reactor core and shielding.\footnote{As a comparison point, we can look at the much larger nuclear-powered variant of the Snark missile. It used the A129 powerplant with a 1.04~m diameter UO$_2$-BeO-fueled ceramic reactor that weighed 1470~kg. The core weighed approximately 1470~kg, and the total powerplant mass was around 4100~kg---approximately 25--30\% of the system weight \cite{comassar1962general} The Snark's planform was radically different from Burevestnik, so the extent to which this is a validating datum is uncertain.} 

\subsection{Propulsor Choice}
The propulsor design (ramjet, turbojet, turbofan, or propfan) provides a key constraint on reactor thermal output, footprint, and system performance. Russia's choice for Burvestnik has not been made public, therefore we evaluate the options here.

\begin{figure}[htbp]
    \begin{center}
    \includegraphics[width=.70\columnwidth]{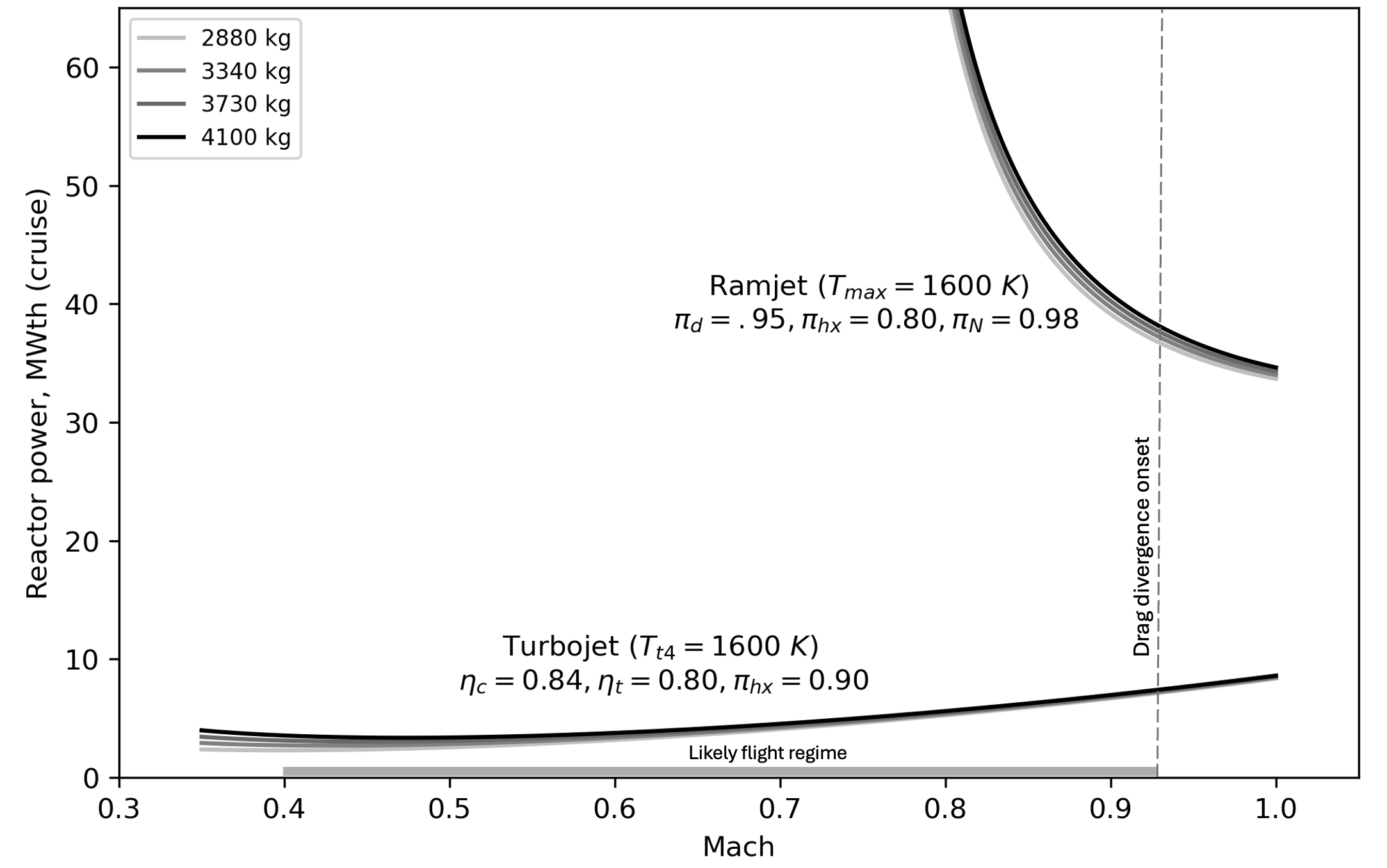 }
    \end{center}
    \vspace*{-5mm} 
    \caption{The thrust required for cruise for a conservative ramjet design is almost a factor of ten larger than for a turbojet. Note that the thermal output of the ramjet required to sustain flight at the observed airspeed ($\approx 0.79~M$) is in excess of 50~MWth. Under these assumptions, 6$^\circ$ climb at sea level as shown in Figure~\ref{fig:PropulsionRequirements} would require in excess of 115~MWth from a ramjet powerplant.}
    \label{fig:RamjetComp}
\end{figure}

Though frequently discussed, a ramjet propulsor is likely infeasible. Ramjets require inlet conditions that generate significant compression and thus are most efficient in the Mach 3--4 range, and generally need at least Mach 1 for reasonable thrust. The planform of Burevestnik, however, limits the craft to subsonic speeds. For a ramjet to fly Burevestnik at subsonic speeds would require a reactor with $>$50~MWth output. Given the available space, that translates to a power density of $>$1kW/cc---a value achieved only in liquid hydrogen-cooled nuclear thermal rockets, exceeding any known air-cooled reactor by a factor of 3--5. Figure~\ref{fig:RamjetComp} illustrates this challenge.

Between turbojet and turbofan options, the use of a fan results in moderate energy-efficiency gains over the turbojet. However, the need for a bypass region for the fan flow reduces the volume available for the heat-exchanger and reactor by a factor of two or more, greatly reducing space efficiency and increasing the reactor-engineering burden. Given that nuclear fuel has extraordinary energy density, but power density is limited by material temperatures, space efficiency rather than energy efficiency is likely to be the deciding factor (the opposite is usually true in hydrocarbon systems). When nuclear powered, it would therefore be more sensible to favor a turbojet. The exception to this might be if the designers chose to use a high-efficiency closed-cycle system to reduce radioactive emissions associated with direct-cycle engines. In such a case, a ducted fan driven by an extremely compact high-temperature inert-gas-cooled reactor would mitigate some material challenges and allow very high energy efficiency (perhaps $>$30\% compared to a turbojet). However, such a complex system has not been demonstrated in ground tests by any country, and the materials and mechanical engineering of ultra-high-temperature systems remain formidable.  We therefore judge that the most likely choice is a direct-cycle turbojet, which is less complex, lighter, and fits better into a narrow fuselage of Burevestnik. Other trade-offs are summarized in Table~\ref{tab:propulsor}. Cycles are discussed further in Section~\ref{sec:reactor} and Table~\ref{tab:cycle}.

\begin{table}[h]
\caption{\label{tab:propulsor}
Properties of various propulsors for endoatmospheric flight.}
\centering
\begin{tabular}{lllcl}
\textbf{Engine cycle}&
\textbf{Complexity}&
\textbf{Mass} &
\textbf{$M_{opt}$} &
\textbf{Limitations} 
\\
\hline
Ramjet & very low & low & 0.9--4~M & Very high minimum speed, poor efficiency\\
Turbojet & moderate & moderate& 0--2.5~M & Moderate energy efficiency  \\
Turbofan & high & moderate & 0--1.5~M & Restricts reactor volume, total power  \\
Closed-cycle fan & very high & moderate & 0--1.0~M & Materials, coolant loss, complex turbomachinery \\
Propfan & high & high& 0--0.9~M & Poor radar cross-section, packaging \\
\end{tabular}
\end{table}
\FloatBarrier

\subsection{Reactor Design}\label{sec:reactor}
This section develops a plausible reactor concept for the Burevestnik by proceeding from overall cycle selection to core configuration, fuel choice, and thermohydraulic performance. Across all engine-cycle calculations, we consider reactor outlet temperatures bounded by the likely limits of candidate structural materials, including SiC, BeO, Rh, and HfC. We also note that the propulsion system may be hybridized, with additional thrust supplied by hydrocarbon combustion during brief high-thrust maneuvers such as launch or terminal obstacle avoidance. Under that interpretation, the reactor need not be sized for peak transient thrust, but rather for the cruise and climb conditions that dominate the vehicle's flight profile (Figure~\ref{fig:DragPolar}). Accordingly, the thermal-power and radiological analyses that follow are based primarily on the expected cruise thrust requirement.

\begin{table}[h]
\caption{\label{tab:cycle}
Properties of various cycles coupling reactors to propulsors. \textquote{Safety} is used here as an inclusive term indicating safety to ground personnel as well as fission product leakage hazards during flight. The mass hierarchy is for small engines ($<$10~MWth), and does not hold for very large systems.}

\centering
\begin{tabular}{lcccccc}
\textbf{Cycle \& Coolant}&
\textbf{Mass}&
\textbf{Exhaust Rad.} &
\textbf{Power Conv.\ Eff.} &
\textbf{Complexity} &
\textbf{Safety} 

\\
\hline
Direct cycle (solid fuel) &+ & - & - & low & very poor  \\
Indirect cycle (liquid fuel) &- & +& + & high & poor  \\
Indirect (molten metal or salt coolant) &- & +& + & moderate & good  \\
Indirect (helium coolant) &- & +& ++ & high & good  \\
\end{tabular}
\end{table}

Given the tight dimensional and mass constraints implied by the observed vehicle geometry, a direct-cycle reactor is the most plausible configuration. The remainder of this section therefore focuses on direct-cycle cores.

\subsubsection{Fuel-System Options for a Direct-Cycle Core}

Within a direct-cycle architecture, fuel form strongly constrains both neutronic feasibility and achievable thermohydraulic performance. Table~\ref{tab:fuels} summarizes the principal fuel classes considered here. These options are evaluated assuming an outer reactor diameter compatible with the observed fuselage width of roughly 85~cm.

\begin{table}[htb]
      \caption{Table of direct cycle fuel options for the Burevestnik turbojet. In referring to criticality limitations, we are assuming a system outer diameter which can fit within the observed Burevestnik fuselage width of around 85~cm. The GE-ANP program tested CERMETS (NiCr+UO$_2$ and BeO--UO$_2$ ceramics), while the NERVA and equivalent Soviet nuclear thermal rocket program tested graphite composite fuels and carbides.}
      \label{tab:fuels}
      \centering
        \begin{tabular}{lll}
        \textbf{Fuel Material}&
        \textbf{Coat/Clad}&
        \textbf{Notes}
        \\
        \hline
        
        SiC-matrix composite	&	None & Small critical dimensions, excellent thermal properties \\
        SiC-matrix particle	&	None & Improved FP retention, but lower fissile loading than composite fuel \\
        CERMET	&	Coating & Criticality limited in assumed geometry w/o exotic matrix material \\
        BeO-UO$_2$ dispersion	&	None & Criticality requires high U fraction, compromising material properties \\
        Graphite composite	&	Coating & Criticality limited in assumed geometry, likely poor durability  \\
        UN, UC, etc pellet &	Clad & Requires cladding, geometry therefore poses heat transfer issues \\
    
        \end{tabular}
\end{table}





\subsubsection{Neutronic and Thermohydraulic Considerations}
For a direct-cycle engine, the dominant geometric tradeoff is the core open fraction. Increasing open fraction permits greater airflow and lowers pressure drop, but it also removes fuel-bearing volume and makes criticality more difficult to achieve. To bound this trade space, we modeled a representative Burevestnik reactor in OpenMC using a right-circular cylindrical core populated by hexagonal fuel elements with cylindrical axial flow channels. This configuration follows the general form of historical air-breathing nuclear propulsion reactors such as HTRE-2, Tory-IIA/C, and XNJ-140, as well as more recent conceptual studies \cite{deng2023conceptual}.

Both fast-spectrum and thermal-spectrum variants were examined. Thermal cases used YH$_{1.5}$ moderation, whereas all designs employed a 10~cm BeO reflector and eight B$_4$C control drums in order to represent an optimistic but plausible upper bound on neutronic performance. Criticality was evaluated with all control elements in the fully withdrawn position. To ensure adequate margin for fuel burnup, we require $k_{eff}=1.05$.

The design variables explored were core diameter, open fraction, fissile isotope ($^{235}$U, $^{233}$U, or $^{239}$Pu), fuel composition (see table~\ref{tab:fuels}). For each neutronic configuration, we use thermo-hydraulic and engine-cycle modeling tools to estimate the optimal flowpath $L/D$ ratio, core pressure drop, specific thrust, and overall thermal efficiency. The propulsion model assumed an 6Al-4V compressor and Inconel-738 turbine, which are common material choices for modern turbomachines. This combined approach identifies the subset of reactor designs capable of satisfying the propulsion requirements established in Section~\ref{sec:requirements}.

\begin{figure}[h!!]
    \begin{center}
    \includegraphics[width=.90\columnwidth]{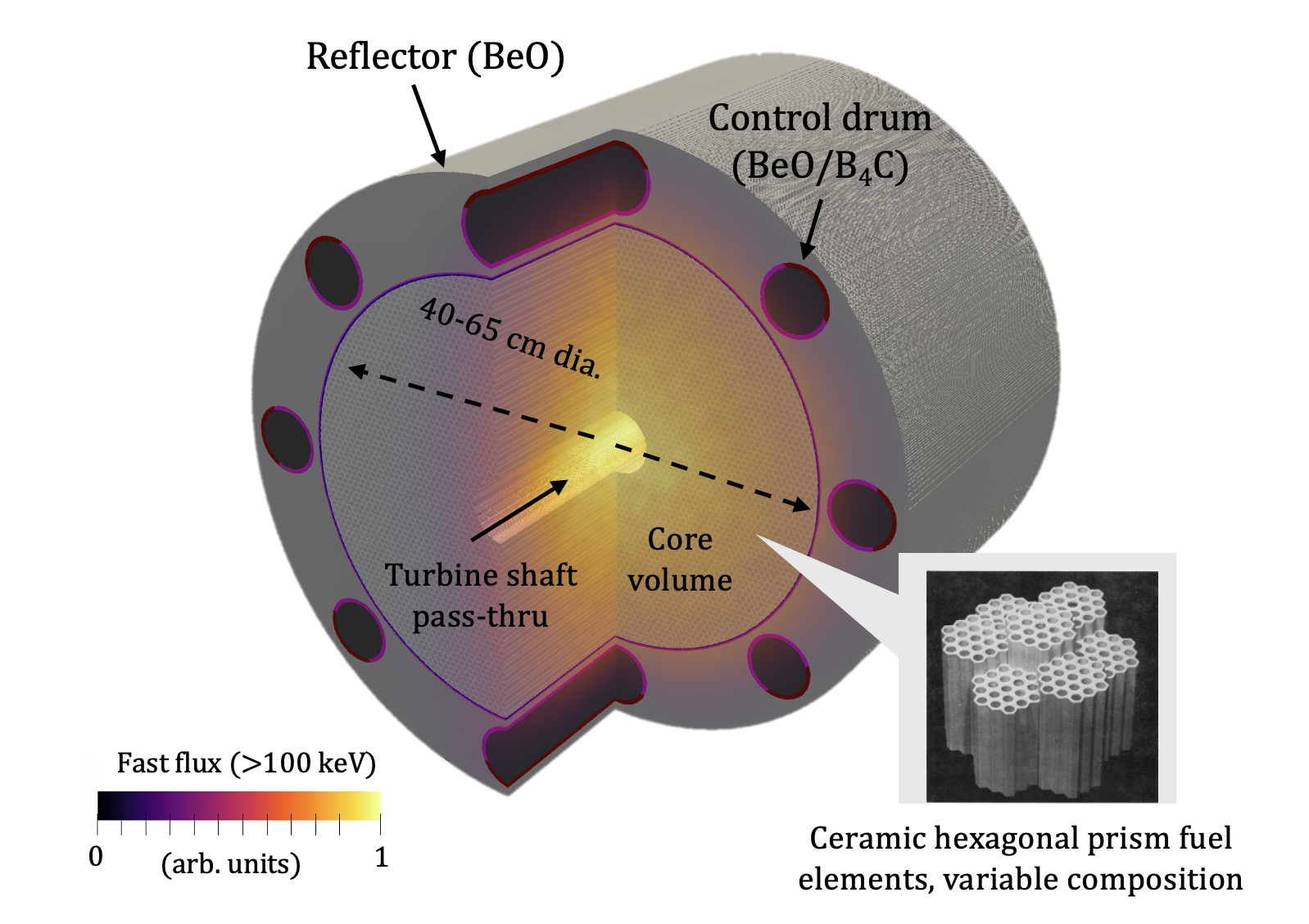 }
    \end{center}
    \vspace*{-8mm} 
    \caption{ Using OpenMC, we modeled the Burevestnik reactor as an inline direct cycle system (similar to the cores used in historical direct-cycle air-breathing propulsion engines). This graphic demonstrates the layout of the engine with a fast-flux overlay (mesh tally) from a $k_{eff}$ search performed as part of parameter-space investigations. Though only a fast-reactor layout is shown here, we explored both fast reactor cores with $SiC-UN$ fuel and thermal reactors using $YH_{1.5}$ moderation. The fast and thermal flux source terms from these Monte Carlo simulations were used to determine shielding and air activation. }
    \label{fig:reactorlayout}
\end{figure}

\begin{figure}[h!!]
    \begin{center}
    \includegraphics[width=.90\columnwidth]{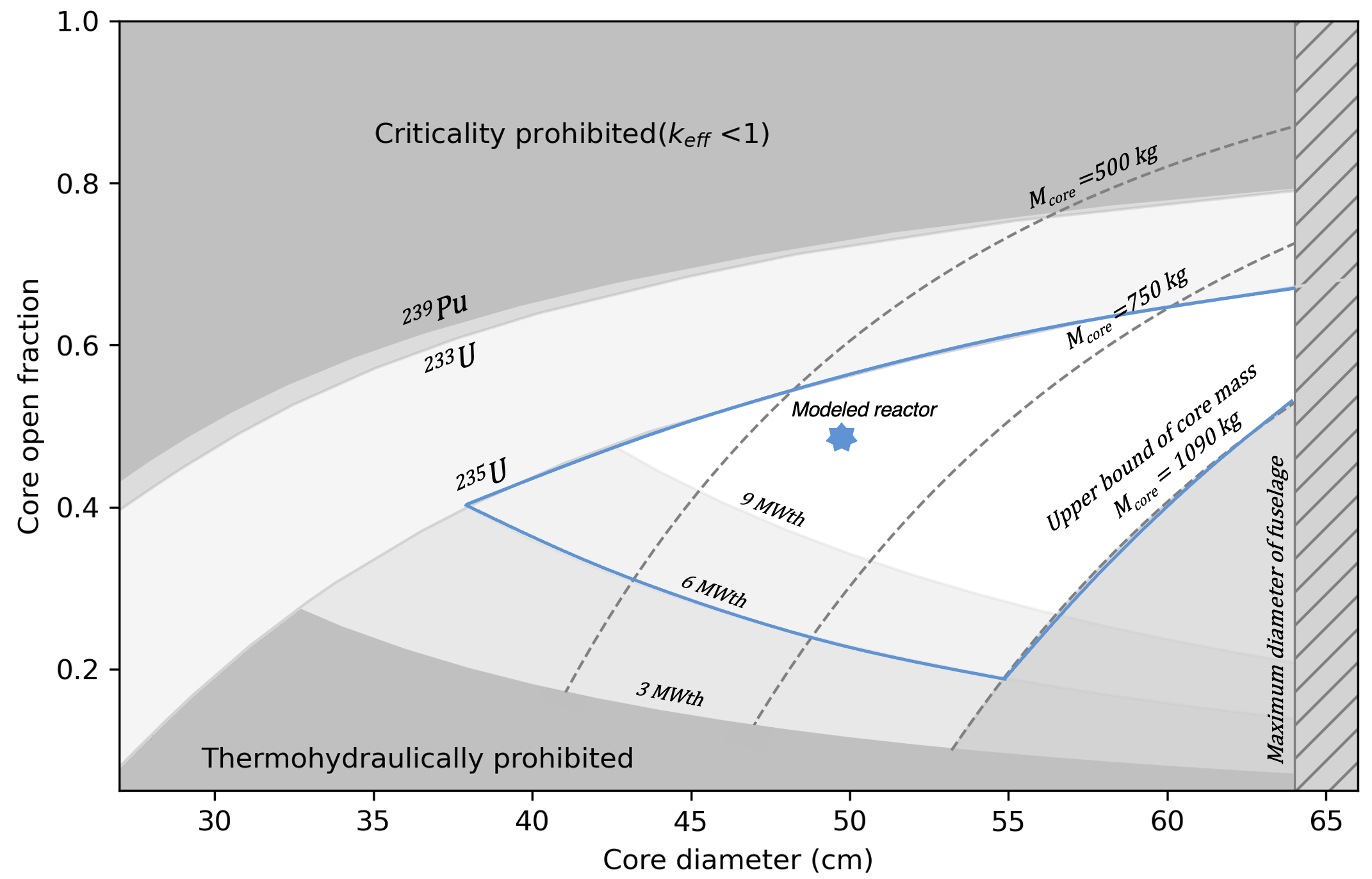 }
    \end{center}
    \vspace*{-8mm} 
    \caption{We bound the design space for a fast-spectrum, direct cycle propulsion reactor using neutronic and thermohydraulic modeling. We have highlighted the likely parameter space for a HEU-fueled Burevestnik core, and plotted the open-fraction/diameter specification of the reactor we use for subsequent modeling tasks. The design space is bounded in the low-open-fraction limit by thermohydraulic considerations. This constraint is set by the choking condition for inviscid flow with heat addition (Rayleigh flow). The curve shown has been drawn for a reactor with an outlet temperature of 1400~K, with a thermal power in intervals of 3, 6, and 9~MWth. These are roughly appropriate for a 3400~kg cruise missile for sea-level cruise, climb, and maneuver respectively. At the high-open-fraction limit, the design space is bounded by criticality limitations for each fissile fuel choice. We model the core in OpenMC as containing hollow hexagonal prisms of SiC-UN or SiC-PuN fuel, with a 10~cm circumferential BeO reflector and eight boron carbide faced control drums. The compressor and turbine have been modeled as homogenized 6AL-4V titanium and Inconel 738, respectively. The OpenMC k-effective search module was used to evaluate the maximum open fraction for each core diameter.} 
    \label{fig:OpenMC}
\end{figure}

We found that fast-spectrum cores are generally superior for this application because they devote more of the available volume to heat-producing fuel rather than moderator. The resulting reference geometry is shown in Figure~\ref{fig:reactorlayout}, and the overall fast-spectrum trade space is summarized in Figure~\ref{fig:OpenMC}.

We found that ANP-style $^{235}$UO$_2$-BeO fuel \textit{cannot} achieve criticality within the volume available. A UO$_2$-BeO dispersion design becomes feasible in this size regime only with $^{233}$U, a fissile material for which Russian stockpiles are believed to be limited. By contrast, a 95\% enriched $^{235}$UN-SiC dispersion fuel can maintain criticality for core masses below 500~kg. This makes SiC-matrix nitride fuel the most credible HEU-based option among the candidates examined.

Other fuel systems, including some CERMET variants, may approach the same lower mass bound, but SiC remains attractive because it combines good manufacturability, favorable neutronics, and oxidation resistance at temperatures above 1400~K. If $^{233}$U or $^{239}$Pu were available, the feasible open fraction would increase further, exceeding 70\% for the SiC-based designs considered here.\footnote{Using $^{233}$U in lieu of $^{235}$U allows critical configurations using 80\%-wt-$^{233}$UO$_2$ in UO$_2$-BeO dispersion fuel. This represents performance somewhat better than a $^{235}$U-SiC core. As UO$_2$-BeO has been used and characterized extensively in air-breathing nuclear propulsion studies dating to the 1950--1960s, a $^{233}$UO$_2$-BeO core could be a safe option from an R\&D risk perspective, though extremely expensive.}

\subsubsection{Reference Reactor for Detailed Analysis}
For the detailed thermohydraulic and engine-performance calculations below, we adopt a representative point near the middle of the feasible design space shown in Figure~\ref{fig:OpenMC}. This reference reactor has a 50~cm core diameter, a 70~cm reflector outer diameter, and a 50\% open fraction. For a $^{235}$U-fueled system this choice lies near the upper open-fraction limit allowed by criticality, while still remaining comfortably within the measured fuselage dimensions.

This reference design is not intended as a reconstruction of the actual reactor, but rather as a plausible and relatively conservative design point within the admissible parameter space. It provides a consistent basis for estimating reactor pressure loss, achievable heat transfer, and overall propulsion performance.



\subsubsection{Heat Transfer and Component Efficiency Assumptions}

For a given reactor design, engine performance depends strongly on heat-exchanger performance, compressor efficiency, turbine efficiency, and pressure loss through the reactor itself.  Although the underlying Brayton cycle is the same as in a combustion-heated turbojet, nuclear-heated systems differ in one important respect: the pressure recovery through the reactor is much poorer than through a conventional combustor. Standard gas-turbine cycle models are therefore not directly applicable, and we instead use a dedicated modeling framework for direct-cycle nuclear turbojets.


Based on a review of compact axial turbomachinery representative of current technology, we adopt isentropic efficiencies of:
\[
\eta_c = 85.4\%\textrm{--}82.8\%, \qquad
\eta_t = 81.4\%\textrm{--}78.8\%
\]
over the pressure-ratio and turbine-inlet-temperature range of interest (1400--1600~K). These values are intended to represent the performance realistically available to a compact missile-scale system.

To model pressure losses in the reactor, we follow Landram's treatment of one-dimensional, steady compressible flow with friction and uniform wall heat flux \cite{landram1997one}. A simple Rayleigh-flow approximation is insufficient because wall friction is significant in the narrow heated channels of a direct-cycle core. Landram expresses the evolution of the squared Mach number, $y=M^2$, as a function of the stretched axial coordinate $\xi = \frac{4fx}{D_h}$:

\begin{gather*}
    \frac{dy}{d \xi} = \left[ {\frac{1+\gamma y}{\frac{1}{\varphi}+\xi}} +\gamma y \right] \frac{y [1+\frac{(\gamma-1)}{2}  y]}{1-y}
\end{gather*}

Where:

\begin{gather*}
    \beta = \frac{(\gamma -1)}{2} M^2, \quad T_o = (1+\beta)T \\
    \varphi = \frac{K}{M_1 (1+\beta_1)}, \quad 
    K =  \frac{4(\gamma-1)}{4\gamma f} \frac{q_w \frac{P_{erH}}{P_{erW}}}{p_1c_1} \\
\end{gather*}
Here the subscript \textquote{1} denotes inlet conditions and \textquote{o} denotes stagnation conditions. The friction factor $f$ is treated as a free parameter and is taken to be approximately 0.01 for ceramic fuel channels in the Reynolds-number regime of interest, consistent with earlier nuclear propulsion studies \cite{layman1962general}. The expected range here is approximately $Re \approx 10{,}000$--$200{,}000$.

    

   


\subsubsection{Flowpath Optimization and Pressure Recovery}

Using this wall-heated flow model, we calculate pressure, velocity, and temperature distributions along a representative reactor channel. For the fixed-open-fraction reference core, the flowpath $L/D$ ratio is then varied to maximize specific thrust at sea-level cruise inlet conditions ($M=0.79$). This procedure is repeated over a range of maximum turbine inlet temperatures to evaluate sensitivity.

The results indicate that specific thrust is only weakly sensitive to the exact reactor pressure recovery factor, with performance peaking near a recovery factor of 0.86--0.88 for the assumed reference design. This corresponds to channel $L/D$ values of approximately 160--180, or physical channel diameters on the order of 3--5~mm for a 70~cm reactor. These dimensions are consistent with existing prismatic fuel-element manufacturing approaches, especially if the fuel is segmented axially.

This result supports two conclusions. First, a reactor within the inferred \textit{Burevestnik} size envelope can transfer the required heat to the airflow at cruise. Second, the fuel geometry needed to achieve that performance is manufacturable using established high-temperature reactor fabrication concepts. Additional details of these calculations are provided in Figure~\ref{fig:ReactorFlow} in Appendix B.

\subsubsection{Resulting Cruise Power Requirement}

The cycle calculations imply that a \textit{Burevestnik}-like nuclear turbojet requires approximately 1200--1300~kWth per kN of thrust at cruise for fuel temperatures in the range 1200--1600~K. Combining this with the thrust requirement inferred from Figure~\ref{fig:PropulsionRequirements} yields an estimated cruise reactor power of
\[
4.3 \pm 1.3~\text{MWth}.
\]

This estimate is based on the median inferred system mass and is intended to characterize the sustained cruise condition, which is the dominant operating point for the vehicle. Higher power would likely be required for climb, altitude changes, or aggressive maneuvering, but those cases may be met in part by hybrid chemical augmentation rather than by increasing reactor design-point power.

\subsubsection{Powerplant Conclusions}

Taken together, the neutronic, thermohydraulic, and engine-cycle results indicate that the most plausible Burevestnik powerplant is a compact direct-cycle nuclear turbojet using a ceramic-fueled, fast-spectrum reactor. Within the available fuselage diameter, this architecture offers the best combination of low system mass, acceptable flow capacity, and achievable criticality; alternative indirect-cycle concepts appear substantially less compatible with the inferred size and power constraints.

For the estimated vehicle mass and drag characteristics, the propulsion system would require about $3.45\pm0.99$~kN of cruise thrust and a mean cruise reactor power of roughly $4.3\pm1.3$~MWth, with significantly higher power needed for climb and maneuver. These results imply that the reactor must be sized to sustain cruise on nuclear heat alone if the system is to achieve effectively unlimited range. However, the projected maneuvering power requirement does not necessarily compel the use of a chemical interburner, since a sufficiently aggressive direct-cycle core may be capable of meeting at least part of that demand without chemical augmentation.

More broadly, our analysis narrows the range of credible propulsion interpretations. In particular, while other layouts such as turbofans or indirect-cycle systems may remain marginally feasible under more optimistic assumptions, a subsonic nuclear ramjet appears highly unlikely. Given the poor efficiency of ramjets in this flight regime and the limited reactor volume available, such a system would require power density and heat-transfer performance well beyond what is plausible for a Burevestnik-sized airframe.

Finally, these design conclusions carry an important radiological implication: the same compactness that favors a direct-cycle turbojet also limits the shielding mass available to suppress air activation and other operational source terms. The inferred propulsion architecture is therefore not only the most plausible from a reactor-design standpoint, but also one consistent with the substantial airborne radiological signature examined in the following section.

\subsection{Powerplant Signatures}

The operation of nuclear propulsion systems within the atmosphere will generate a radioisotopic trace from three sources: activated air, activated structural material released by erosion, and---for direct-cycle systems---fission products escaping from the fuel. In this paper, we quantify only the first of these. Fission-product release is likely important, especially for noble gases, but depends strongly on fuel form, temperature, and geometry.\footnote{In any case, it is highly likely that the bulk of the noble-gas fission product inventory will leak during operation.} Likewise, estimating signatures from eroded structural materials would require design details that are not presently available. The analysis below therefore focuses on activation products generated in air.

To estimate this source term, we modified the OpenMC reactor models developed for the neutronic design study. We adopt the same representative reactor geometry used in the preceding section: a 50~cm core diameter with 50\% open fraction. We examine both fast-spectrum (SiC-UN) and thermal-spectrum (YH$_x$-moderated UN) reactors, each in direct- and indirect-cycle form, giving four total cases. The direct-cycle models retain the full core-flow geometry described earlier; the indirect-cycle models omit the compressor and turbine materials and do not model a coolant. In the direct-cycle cases, the core flow is represented as atmospheric air at 1400~K and 15~atm.

Each model is surrounded by a variable mass of $^6$LiH treated as idealized $4\pi$ neutron shielding, spanning 0--1000~kg, and then by a 10~km spherical volume of STP air to capture neutron leakage into the atmosphere. We tally the principal activation reactions---including $(n,\gamma)$, $(n,2n)$, and $(n,p)$---in both the surrounding air and, for direct-cycle cases, the air passing through the core. The resulting isotope production rates are listed in Table~\ref{tab:capture}.



 \begin{table}[!h]
 \caption{\label{tab:capture}%
OpenMC simulations allow the estimation of activation products resulting from nuclear flight. This table presents those with half-lives long enough to act as viable detection signatures, with values are given as a function of $^6LiH$ shielding mass surrounding the reactor in $4\pi$. All values are reported as Bq of product produced per MW-hr. The \textquote{core flow} value represents production in the air flowing through the reactor core in a direct cycle design at 15~atm and 1400~K. We have excluded very short half life and low total activity isotopes unlikely to be detected. \textit{Note: Due to cross-section inconsistencies, we have excluded $^{13}$N and $^{35}$S.} }
   \centering
   \begin{NiceTabular}{|c|l|l|l|l|l|l|l|l|l|}
     \hline
     
      &  & \textbf{$^{14}C$} & \textbf{$^{37}Ar$}& \textbf{$^{39}Ar$} & \textbf{$^{41}Ar$} & \textbf{$^{79m}Kr$} & \textbf{$^{83m2}Kr$} & \textbf{$^{85m}Kr$} & \textbf{$^{125}Xe$} \\ \hline
      & \textbf{Shielding} & \textit{5730~y} & \textit{35~d}& \textit{303~y} & \textit{1.81~h} & \textit{50~s}  & \textit{1.83~h} & \textit{4.48~h} & \textit{16.9~h} \\ \hline
      \Block{5-1}{\rotate{\textbf{Thermal}}}
      &None	&	2.86E+08	&	1.76E+09	&	1.67E+04&	3.15E+13	&	1.24E+10 &	9.62E+10	&	2.69E+09	&	1.45E+08\\ 

      &250~kg	&	6.73E+07	&	4.11E+08	&	5.41E+03	&	7.33E+12	&	2.95E+09	&	2.25E+10	&	6.48E+08	&	3.38E+07\\
      
      &100~kg	&	9.55E+07	&	5.74E+08	&	3.86E+03 &	1.02E+13	&	4.40E+09	&	3.36E+10	&	1.00E+09	&	4.88E+07\\
      
      &500~kg	&	5.47E+07	&	3.35E+08	&	3.15E+03&	5.97E+12	&	2.36E+09	&	1.80E+10	&	5.09E+08	&	2.71E+07	\cr
      
      &1000~kg	&	5.21E+07	&	3.20E+08	&	3.01E+03	&	5.69E+12	&	2.21E+09	&	1.71E+10	&	4.88E+08	&	2.57E+07\\
      
      &Core flow &	4.80E+05	&	2.36E+06	&	2.23E+01	&	4.25E+10	&	1.92E+07	&	1.14E+08	&	4.68E+06	&	1.55E+05\\

      \hline \hline

      \rowcolor{aliceblue}
      \Block{5-1}{\rotate{\textbf{Fast}}}
      &None &	4.29E+08	&	2.64E+09	&	2.48E+04&	4.70E+13	&	2.01E+10	&	1.54E+11	&	4.52E+09	&	2.26E+08\\
      \rowcolor{aliceblue}
      & 100~kg	&	1.65E+08	&	1.04E+09	&	9.82E+03	&	1.86E+13	&	8.19E+09	&	6.20E+10	&	1.85E+09	&	9.02E+07\\
      \rowcolor{aliceblue}
      &250~kg	&	9.87E+07	&	5.99E+08	&	5.63E+03&	1.07E+13	&	4.76E+09	&	3.56E+10	&	1.07E+09	&	5.17E+07 \\
      \rowcolor{aliceblue}
      & 500~kg &	8.02E+07	&	4.88E+08	&	4.59E+03&	8.71E+12	&	3.86E+09	&	2.89E+10	&	8.75E+08	&	4.19E+07	\\
      \rowcolor{aliceblue}
      &1000~kg &	3.89E+07	&	2.35E+08	&	2.21E+03	&	4.20E+12	&	1.88E+09	&	1.40E+10	&	4.22E+08	&	2.03E+07 \\
      \rowcolor{aliceblue}
      &Core flow &	2.15E+05	&	1.13E+05 & 1.04E+00	&	4.23E+09	&	4.84E+06	&	3.81E+06	&	1.51E+06	&	4.76E+03 \\ \hline
      
   \end{NiceTabular}
 \end{table} 
 
These simulations indicate that both direct- and indirect-cycle reactors would produce a detectable atmospheric activation signature at MW-scale power. Across the shielding range considered, total activation exceeds roughly 5~TBq per MW-hr in all cases. The dominant short-term signature is $^{41}$Ar, which accounts for at least 95\% of the activity one hour after production and should remain detectable for many hours. Longer-lived products such as $^{85m}$Kr and $^{14}$C are produced in smaller quantities but may also be useful tracers.

The fast-spectrum design generally produces the larger total activation signal in the surrounding air. However, activation of the direct-cycle core flow is lower for the fast reactor because high-energy neutron capture cross sections are smaller. Although we did not explicitly tally the spatial distribution of production, the simulations indicate that most activation occurs within roughly a kilometer of the vehicle. Overall, these results suggest that atmospheric air activation alone would provide a substantial and persistent signature for a Burevestnik-like nuclear propulsion system.


\section{Strategic Implications}

Open-source information about the basing, readiness posture, command-and-control plans, and intended strategic purposes are extremely limited. We can speculate as to possible strategic implications based on the physical characteristics of the system. Most notably, as a nuclear-powered system, it has the potential for extreme flight duration and thus range. This could allow the missile to travel along circuitous flight paths to help evade radar and make missile interception difficult. Long duration flight could also allow for early launch during periods of strategic warning, reducing or eliminating counterforce options of Russia's adversaries.  The same could also be framed as a delayed-attack option, such as a dead-hand-like system in the event of leadership decapitation. These are all speculative. 

What Burevestnik does not do is obviously transform Russia's ability to conduct nuclear strikes or overcome its deterred state by giving Russia a decisive advantage.  Russia already fields a large and diverse strategic nuclear force that includes intercontinental ballistic missiles, submarine-launched ballistic missiles, and . Despite the recent degradation of Russia's strategic arsenal, existing and near-future missiles, bombers, and both air- and sea-launched cruise missiles.  Even though some of these systems have seen stress or degradation in recent years, and even though the United States has deployed increasing missile defenses, experts do not assess that there is an imminent collapse in Russia's ability to inflict a deterring second strike.\cite{https://sgp.fas.org/crs/natsec/IF10541.pdf} 

The rationale for developing Burevestnik is thus twofold. First, hedging against future missile defenses by improving on what Russia already achieves with traditional cruise missiles, namely approaching from unexpected azimuths, terrain hugging for long periods of time, and other strategies that complicate tracking but are fuel inefficient. The nuclear propulsion system allows Burevestnik to implement these strategies more comprehensively. But this should not be overstated: the Burevestnik does not appear to incorporate any novel design features which decrease its detectability below that of conventionally-powered cruise missile in the field of view of a radar system. Indeed, its large size may create additional signal. Because it appears to rely on solid rocket boosters, its launch may be visible to space-based infrared sensors, which can queue a variety of tracking modalities such as radar moving-target indication. If Burevestnik is detected and tracked by some means, the subsonic flight envelope means Burevestnik is subject to interception. 

The second logic appears to be psychological. Burevestnik descends from a long-line of ``prestige weapons,'' including Avangard and Poseidon. These systems are build on the legacy of psychological deterrence going back to Tsar Bomba. Because they provide genuinely novel effects, they can instill fear in the adversary---even if those capabilities are not decisive or practical. 

The extent to which Burevestinik will manifest these gains is an open question.  While it is true that reactors provide enormous fuel density, they do not enable truly indefinite flight. Reactors will eventually burn enough fuel to lose reactivity, and materials will eventually degrade to limiting conditions---likely long before the fuel is exhausted. In particular, radiation damage such as radiation-assisted creep \cite{whitmarsh1971neutronic}, embrittlement, or void formation can occur to structural materials in the core, engine turbine, and airframe. These phenomena will limit the mechanical lifetime of the missile. Radiation damage to electronics may also limit the functional life of command and control systems. Finally, the exhaustion of consumables (e.g., lubricants, chemical fuels), including the damage of these consumables by radiation, is also a consideration.  We still assess the vehicle is likely able to fly for tens to hundreds of hours, sufficient for basic radar-evasion tactics, but these limitations will curtail loiter-based tactics.

The strategic-stability consequences of long-duration flight are mixed. To the extent that such a capability improves Russia's ability to penetrate missile defenses and preserves its retaliatory strike, the weapon can be argued to be stabilizing within theories that treat mutual vulnerability or assured second-strike capability as central to deterrence. At the same time, a weapon whose flight path and arrival time are unknown may aggravate crisis instability by increasing fears of surprise attack, encouraging escalation in the face of uncertainty. If Burevestnik were launched early as a survivability measure---much like the dispersal of road-mobile missiles and submarines---such an action could serve as a clear deterrent signal to check escalation, but it could also heighten tensions and be interpreted as preparation for attack. On balance, the weapon’s effects on stability appear ambiguous rather than clearly stabilizing or destabilizing.

Where Burevestnik presents a definite challenge is in testing and fielding of the system without an intention to attack. As shown, the operation of a direct-cycle nuclear propulsion system within the atmosphere will create sizable radioactive source terms resulting from air activation, erosion of activated structural materials, and diffusion of fission products out of the fuel. Even under the assumption of complete fuel encapsulation, operation and testing near human habitation poses hazards from airborne radionuclides. The risk of a crash is also significant---as demonstrated by the August 2019 Nyonoksa accident---with dramatic and long-lasting human health impacts. These risks frame Russia as an irresponsible and dangerous actor even during peacetime, and they create incentives for population-protective interception of test articles that could, in turn, generate unintended escalation risk. 

While the Burevestnik is the only nuclear aircraft claimed to be operational, it is unlikely to represent the totality of defense air-breathing nuclear propulsion R\&D in Russia. Ultra-long-duration flight enables missions unachievable with conventional propulsion \cite{comassar1962general}, \cite{cheng2024novel}. Beyond strike roles (as in the Burevestnik), missions in intelligence, surveillance and reconnaissance (ISR) may also benefit from such extreme persistence. This may enable the deployment of advanced radars, communication relays, or other high-power-requirement payloads for weeks or months without refueling. Given the proliferation of antisatellite capabilities, indefinite-duration systems may serve a useful role as \textit{pseudo-satellites,} a way to rapidly deploy functional replacements for space-based capabilities that could be degraded or lost early in a conflict. Burevestnik is likely to be seen as a stepping stone to broader aerospace nuclear capabilities.


\section{Conclusions}


Based on planform, we assess that Burevestnik flies at high-subsonic speed. Aerodynamic and engine cycle modeling suggests that the Burevestnik most likely uses an inline, direct-cycle nuclear turbojet with a turbine-inlet temperature in the range of 1200--1600~K. However, we cannot conclusively state whether the system is direct or indirect cycle, or narrow down fuel choices between CERMETs or ceramics. Based on heat transfer and flight regime limitations, it is almost certainly not a ramjet. In this configuration, heat would be supplied by a compact reactor with a core diameter of around 50~cm. At cruise, we estimate achievable reactor thermal power of $4.3\pm1.3~MWth$, with peak power demand during climbing and terminal maneuvering exceeding $15~MWth$, which may require chemical thrust augmentation. In all designs, we find that neutrons escaping the reactor and being captured in air will generate a significant radioactive trace, at least 5~TBq per MW-hr of activation products alone (exclusive of any leaked fission products or activated structural materials). 

Taken as a whole, Burevestnik does not appear to be a strategic game-changer, nor is there evidence that it gives Russia a decisive new warfighting advantage beyond the retaliatory capability provided by its existing nuclear forces. Its most plausible significance lies instead in the margins: as a hedge against future missile defenses, a psychologically potent ``prestige weapon,'' and a potentially useful niche capability for complicating warning, tracking, and defense planning. Those effects are real enough to merit attention, but they do not obviously overturn the basic structure of deterrence or strategic stability, and its net consequences remain ambiguous. The most serious issue may be that nuclear-powered cruise missile introduces unusual peacetime risks arising from environmental and political hazards associated with testing, accidents, and recovery of a flying nuclear reactor with no containment. In that sense, Burevestnik is best understood not as a revolutionary weapon, but as a costly, hazardous, and symbolically charged system with the potential to kick off a broader aerospace nuclear-propulsion race.



\appendix

\newpage

\section{Drag Workup Validation}\label{app:delta-verify}
\begin{figure}[htbp!]
    \begin{center}
    \includegraphics[width=1.0\columnwidth]{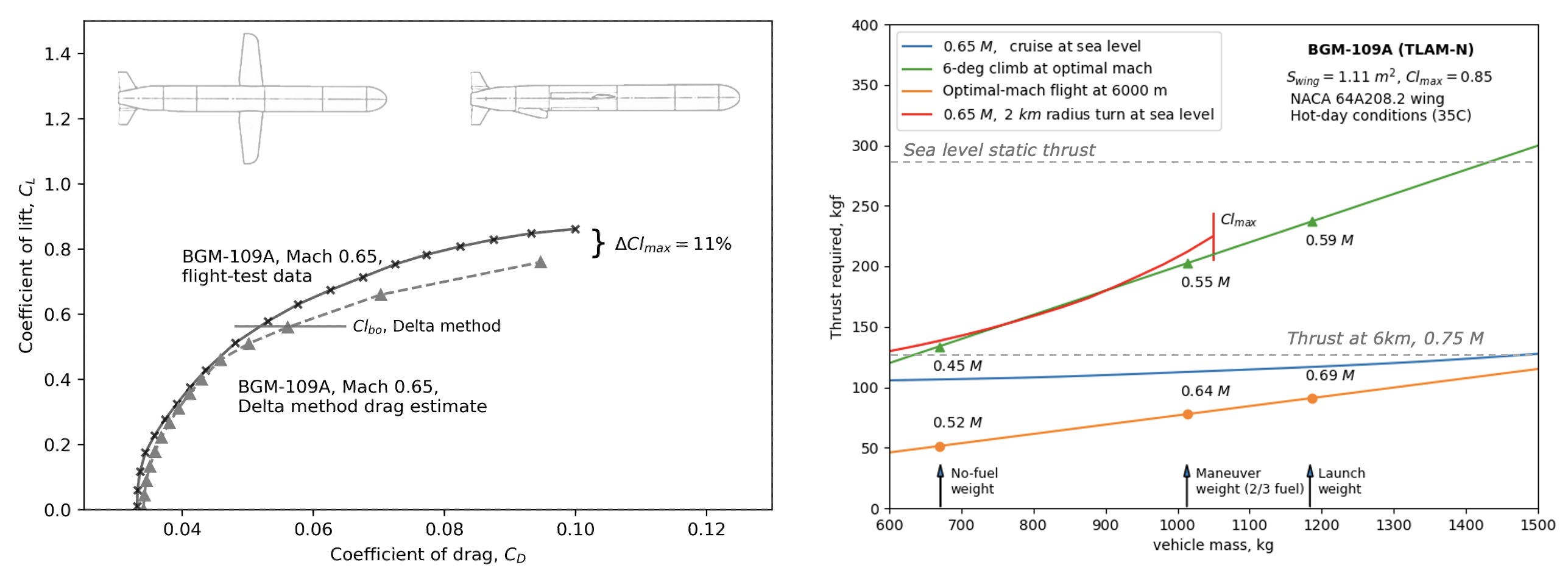 }
    \end{center}
    \vspace*{-2mm} 
    \caption{ \textbf{Left:} To establish the consistency of the Delta method when applied to modern cruise missiles, we benchmarked our procedure and code using the published AGM-109A (equivalent to BGM-109A/TLAM-N) drag polar from Kuchta et al \cite{kuchta1981technology}. As can be seen in the figure, the calculated and measured drag polars are highly consistent before buffet onset (horizontal line). \textbf{Right:} Using the drag polar and knowledge of the system mass, we can extract thrust requirements for the maneuvers identified in Feifel and Kerkam. This has been performed as a benchmarking exercise for the TLAM-N to verify the applicability of the method to Burevestnik. As can be seen in the plot, the missile has sufficient thrust for the mission, but is only capable of performing a steep terrain avoidance maneuver with less than 2/3 fuel load. } 
    \label{fig:DragPolar-BGM}
\end{figure}

We benchmarked the Delta Method for cruise missiles by applying this drag buildup procedure to the AGM-109A / BGM-109A / TLAM airframe using published data on the planform, airfoil, mass and propulsion system \cite{craig1981flight}. As shown in \ref{fig:DragPolar-BGM}, the agreement is excellent between flight measurements and the Delta Method, which tends slightly conservative (underestimates $C_L/C_D$). To provide an additional benchmark using a heavy cruise missile using a supercritical airfoil, we repeated this workup procedure for the Boeing Long Range Conventional Strike Weapon (LRCSW) entrant using the wing, body and airfoil data presented in Feifel and Kerkam \cite{feifel1992propulsion}. In this case, the method slightly under-predicts zero-lift drag at design Mach by around 6\% ( $C_{D_{min}-Delta} = 0.0382,\, C_{D_{min}-meas} = 0.0401$), but accurately captures the shape of the drag polar to high $C_L$ values.


\begin{table}[!htpb]
    \caption{Drag workup inputs for the TLAM from Craig \protect\cite{craig1981flight} and Lewis \protect\cite{lewis1992long} (left), and Burevestnik (right). Note that for the Burevestnik, $S_{wet-body}$ has been estimated from our CAD model, and may vary based on empennage design. We establish an upper limit by assuming the maximal case in which the empennage has the same cross-section as the canister interior, and does not taper.}
    \begin{minipage}[t]{0.45\linewidth}
      \centering
        \begin{tabular}{lr}
        \textbf{Factor}&
        \textbf{Value (TLAM)}
        \\
        \hline
        
        $S_{wet-wing}$	& {2.23}~m$^2$ \\
        $S_{wet-fuse}$	& {9.12}~m$^2$ \\
        $S_{wet-tail}$	& {0.84}~m$^2$ \\
        $AR_{wing}$	& {5.74}  \\
        Fineness ratio	& {9.73}  \\
        Wingspan	&	2.59~m\\
        Wing mean chord &	0.43 \\
        Sweep angle	&	0$^\circ$  \\
        Wing, tail T/C	&	8.2$~\%$  \\
        Vert. stab. length	&	0.42~m \\
        Horz/Vert stab mean chord	&	0.25~m\\
        Horz/Vert. stab sweep	&	0$^\circ$ \\
        Vert. stab. length	&	0.42~m \\
        
        \end{tabular}
    \end{minipage}%
    \begin{minipage}[t]{.45\linewidth}
      \centering
        \begin{tabular}{lr}
        \textbf{Factor}&
        \textbf{Value (Burevestnik)}
        \\
        \hline
        $S_{wet-wing}$	& {5.59} \pm {0.83}~m$^2$ \\
        $S_{wet-fuse}$	& {21.40} \pm {4.3}~m$^2$ \\
        $S_{wet-tail}$	& {2.22} \pm	{0.39}~m$^2$ \\
        $AR_{wing}$	& {9.25} \pm {.56} \\
        Fineness ratio	& {10.79} \pm {1.9} \\
        Wingspan	&	5.60	\pm	0.18~m\\
        Wing mean chord &	0.595		\pm	0.02~m \\
        Sweep angle	&	14.8 \pm	3.8$^\circ$ \\
        Wing T/C	&	6.46  \pm .015~\%  \\
        Vert. stab. length	&	1.18	\pm	0.16~m  \\
        Horz/Vert stab mean chord	&	0.393	\pm 0.16~m \\
        Horz/Vert. stab sweep	&	18.6	\pm $3.5^\circ$ \\
        Horz. stab span	&	2.38	\pm 0.24~m \\
        \end{tabular}
    \end{minipage} 
\end{table}







\section{Measurement of Airspeed using SRB Exhaust Disturbances}
\label{app:exhaust}
\begin{figure}[htbp!]
    \begin{center}
    \includegraphics[width=.86\columnwidth]{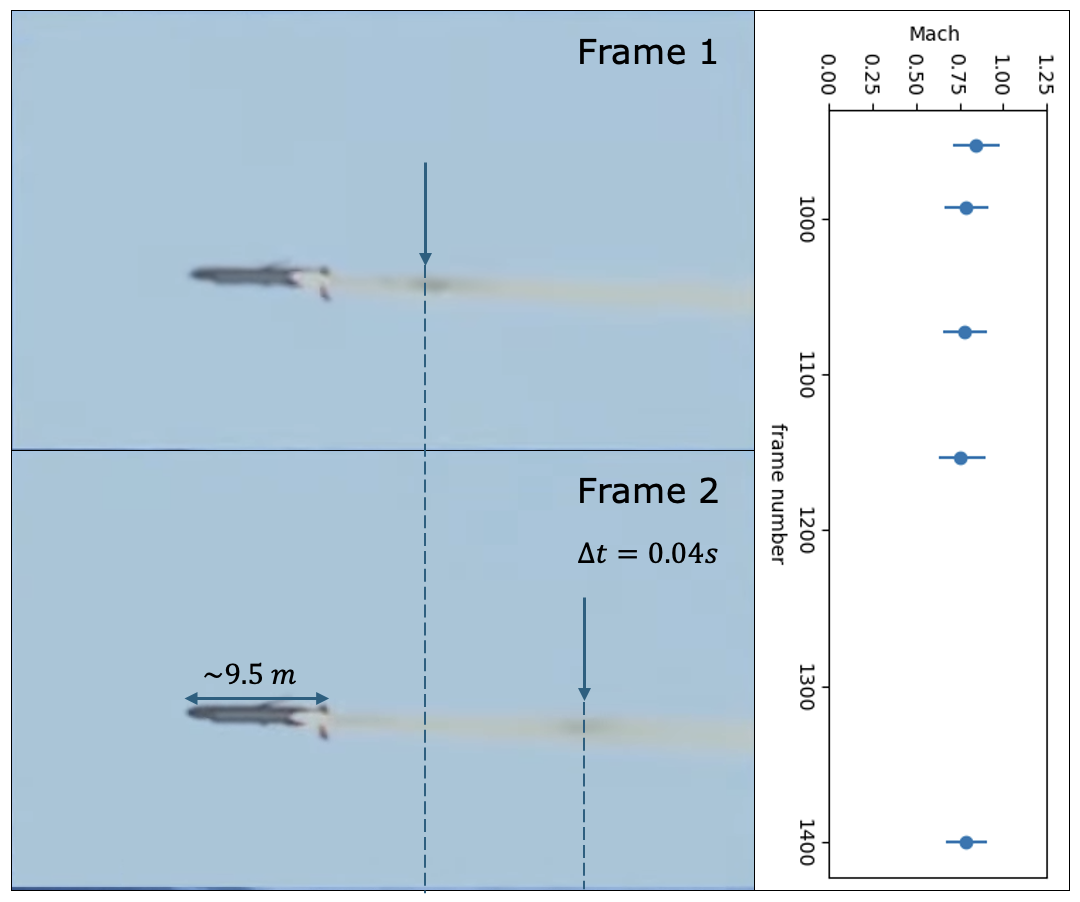 }
    \end{center}
    \vspace*{-5mm} 
    \caption{Exhaust disturbances observed in the solid-rocket booster plume have been tracked across multiple frames to yield estimates of airspeed. Frames 953 and 954 are shown here for illustrative purposes, and airspeed estimates are displayed with respect to frame number at right.}
    \label{fig:FlightMeasurements}
\end{figure}

During the flight video, frame-by-frame viewing shows irregularities in the solid rocket booster exhaust which appear as puffs of black vapor. Literature suggests that for most solid rocket boosters, the plume is close to fully mixed and in equilibrium with the surrounding atmosphere after 50-80 nozzle diameters. Under the assumption of a 10~cm nozzle \footnote{This cannot be easily measured in the images available, and is based on comparison to similar solid rocket boosters in the 500~kg thrust class}, the exhaust should be static with respect to the surrounding air by the time it is approximately 5-8~m behind the aircraft. Prior work has established the length of the missile as approximately 9.5~m. Therefore, by tracking the movement of these puffs from frame to frame (once they are beyond this mixing distance), and using the fuselage as a ruler, we can estimate the airspeed of the missile. Using this analysis technique on the best series of frames (953--955), we arrive at an airspeed estimate of $0.79 \pm 0.13~M$, as shown in \ref{fig:FlightMeasurements}. Averaging subsequent measurements in which we only have two data points, we arrive at an airspeed estimate of $0.83 \pm 0.28~M$. These values are entirely consistent with the missile planform and phase of flight shown in the video. It should be noted that we are performing these estimates assuming no video-speed manipulation, and that the plume is fully in equilibrium during the measured frames.

\section{Reactor Neutronic and Thermohydraulic Modeling Assumptions}
\label{app:reactor}
Based on airframe measurements, we estimate a reactor core of no more than 64~cm in diameter, assuming a 10~cm thick beryllium oxide external reflector. This yields a total outside diameter of 84~cm, without accounting for the engine casing. The goal of this modeling is to establish the limits of the reactor parameter space: our OpenMC simulations determine the upper limit of the core \textquote{open fraction,} \footnote{Which we define as the volume of the air flowpaths divided by the core volume} which is constrained by the need to create a critical assembly. The open fraction is limited on the lower edge by thermohydraulic limits imposed by flow choking. Our OpenMC simulations assume a simplified cylindrical core design with external boron carbide-faced control drums and a fixed-thickness reflector. The axial pass-through for the turboshaft has been fixed as a 7.5~cm Be cylinder, as beryllium has excellent nuclear and mechanical properties, and has been used in prototype gas turbines \cite{corrigan1964applications}. We have modeled the compressor as a 6Al-4V titanium cylinder in front of the reactor, which has been homogenized to a representative density to approximate the length of a typical compressor ($\approx 40~cm$). The turbine is modeled as a similarly homogenized Inconel 738 cylinder behind the reactor. 

As little information is available on the reactor fuel design, we bound the design space using two designs: a fast reactor using UN (or PuN)-SiC \textquote{CER-CER} fuel, and a thermal reactor design using yttrium hydride ($YH_{1.8}$) moderator and UN (or PuN) fuel dispersed in SiC \footnote{To determine the optimal moderator to fuel ratio in the thermal design, an additional iterative criticality search step was performed for each reactor size and open fraction increment.}. In both cases, we model the core as consisting of hexagonal fuel elements with cylindrical flow paths composed of homogenized fuel, SiC matrix and moderator (if applicable). This perforated hexagonal element design follows HTRE-2, XNJ-140, TORY-IIA, TORY-IIC, and the designs proposed by Wen\cite{wen2023numerical} and Lu\cite{lu2023numerical}. Based on a literature review of CER-CER fuels, the fuel:matrix volume ratio has been taken as 1:1 throughout this investigation for both fast and thermal designs. Note that fissile loadings of $^{235}$U,$^{233}$U, and $^{239}$Pu have been used in Figure~\ref{fig:OpenMC} and that all models use a reactor geometry of a right equilateral cylinder. We have selected these fuel and matrix materials based on maximizing neutronic and thermal performance. UN and PuN offer some of the highest fissile material densities for materials capable of high temperatures. While other fuel and matrix/cladding material selections are possible, this material selection is highly neutronically efficient, and likely to be very near the lower limit for core size using currently-known materials. 


To estimate a reasonable range of values for the core open fraction, we have used OpenMC to iterate the open fraction at each reactor core diameter increment using the criticality search capability. In evaluating the limit on this parameter, we use a threshold of $k_{eff}=1.05$, as the use of structural materials in the core, sensors and other engineering practicalities will erode the reactivity margin. All core components have been modeled at 1600~K, and the core air flow has been taken as dry 1600~K air at 15~atm. The reflector and central reflector/turbine shaft element have been modeled at 800~K.

OpenMC modeling showed that the fast/intermediate reactor design had a considerably higher maximum open fraction than the thermal reactor design at all core sizes. For a $^{235}$U-fueled fast reactor using UN in a SiC matrix (50\% UN by volume), the maximum open fraction is 52.5\% for a 50~cm core diameter. Using $^{233}$U, the same 50\% volume fast UN-SiC system is critical to 68.0\% open fraction. The $^{239}$Pu-nitride-fueled core allows a 70.4\% open fraction at the same diameter. While these open fractions are physically possible from a criticality standpoint, it should be noted that open fraction is also subject to manufacturing constraints. The TORY reactor program found that an open fraction of greater than 50\% was not possible with their optimized flow path diameter of  0.249~in (6.32~mm). For flow paths in the near-optimal range for this reactor size (3-5~mm), manufacturing and structural constraints make open fractions above 60\% likely unfeasible. As our models show that the fast reactor designs allow higher open fractions, the core diameter versus open fraction curves plotted in Figure~\ref{fig:OpenMC} use the fast reactor model \textit{only.}  

One alternative design has been investigated, using BeO-UO$_2$ dispersion fuel with $^{233}$U fissile loading. This design retires most of the material science risks, as $BeO$ dispersion fuel with up to 80\% UO$_2$ by mass was tested in the ANP program with considerable success. In our simulations, this design shows open-fraction performance intermediate between the $^{235}$UN-SiC core $^{233}$UN-SiC cores. However, like all  $^{233}$U cores, the cost of producing fissile material without excessive $^{232}$U would be exceptionally high. A HEU-fueled version would not be critical with a reasonable open fraction until a core diameter of greater than 70~cm, ruling it out as an option for the Burevestnik.

In reactor air activation simulations, we have used the SiC-matrix core models described above, though with a geometry fixed at 50~cm diameter and 50\% open fraction. As the leakage fraction varies by a few percent between the upper and lower limits of the reactor design parameter space, we do not anticipate that this reactor geometry choice will modify the results significantly.

\begin{figure}[h!!]
    \begin{center}
    \includegraphics[width=1.0\columnwidth]{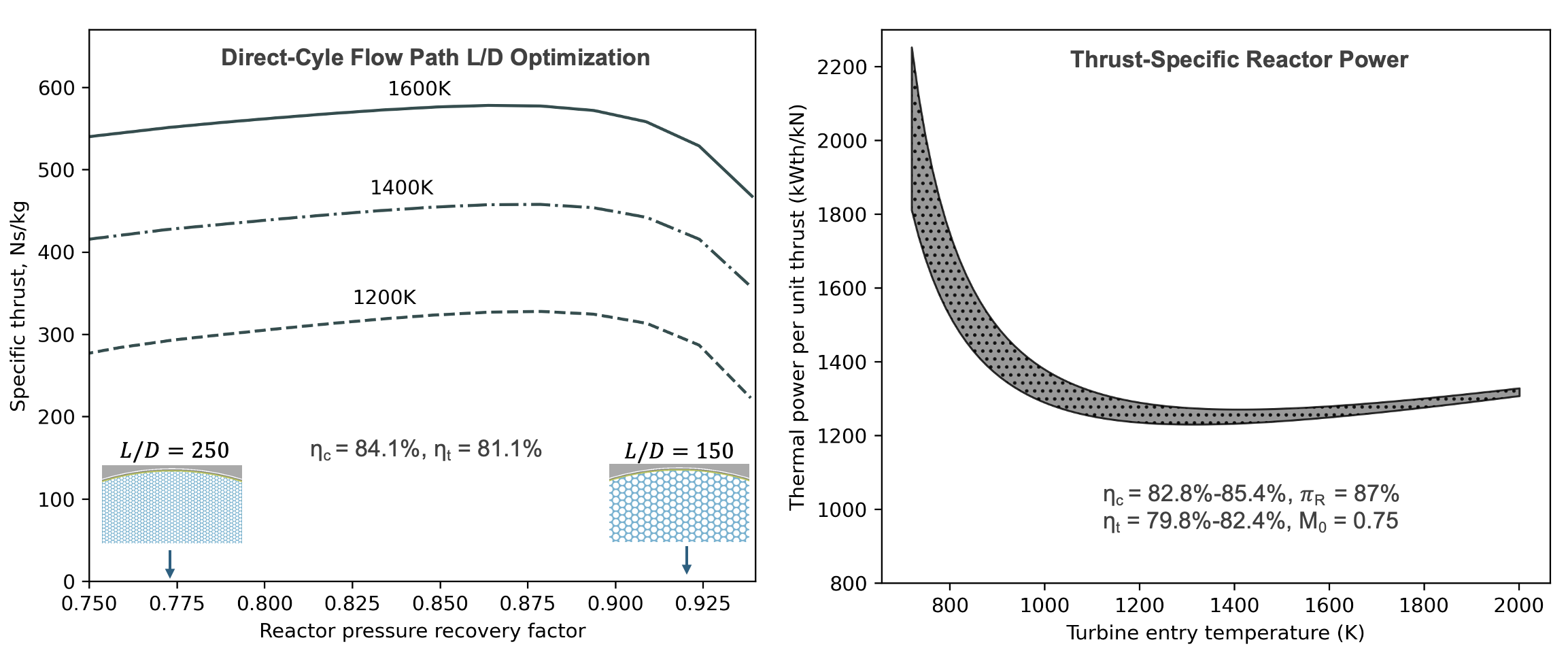 }
    \end{center}
    \vspace*{-2mm} 
    \caption{  \textbf{Left:} By varying reactor flowpath L/D, the influence of the reactor pressure recovery factor, $\pi_r$, can be studied with respect to specific thrust. To simplify analysis, we have fixed the turbine and compressor efficiencies at median values of the left figure. The engine’s thrust per unit flow is highly sensitive to turbine entry temperature, and has limited sensitivity to pressure recovery factor. The nozzle diameter assumed here is 35~cm, within the range of possible values given observations of the Burevestnik.\textbf{Right:} The conversion of thermal energy to thrust in combustion-powered turbomachinery is generally characterized by thrust-specific fuel consumption (TSFC). For a nuclear propulsion system, this metric is not meaningful, and can instead by stated as thrust per unit of thermal power (kWth/kN). As shown here, the efficiency of this conversion is highly sensitive to turbine entry temperature and weakly sensitive to compressor and turbine efficiencies. Based on reactor material constraints, we anticipate a turbine entry temperature of several hundred kelvin lower than the maximum fuel centerline temperature.  
} 
    \label{fig:ReactorFlow}
\end{figure}

We have also conducted thermohydraulic analysis to determine the optimal flowpath geometry for a fixed open fraction. Using Landram's treatment of wall-heated flow as detailed in 3.5.2, we evaluate the aspect ratio of channels with respect to pressure drop and overall cycle efficiency, as shown in Figure \ref{fig:ReactorFlow}. For these calculations, we fix the maximum fuel surface temperature to the maximum turbine entry temperature (1200--1600~K) \footnote{As the fuel elements are relatively small, the difference between fuel centerline and surface temperature is relatively small. Omitting this part of the analysis has a small impact on the result}.

These calculations of optimal heat-exchanger pressure recovery can be used to estimate the efficiency of the Burevestnik powerplant in converting heat to thrust. As the metric of thrust-specific fuel consumption (TSFC) is not meaningful for nuclear powered systems, we define an thrust efficiency metric of kWth/kN. The results of this analysis are shown in \ref{fig:ReactorFlow}. In the analysis of thrust efficiency, we have fixed the reactor pressure recovery to the iteratively determined optimal value, and used cycle modeling tools to calculate the thrust efficiency metric with respect to turbine inlet temperature. We bound the performance envelope by using the upper and lower limits of turbine and compressor efficiencies identified in our prior literature review.

\clearpage
\FloatBarrier
\bibliography{references}
\end{document}